%% file: main.tex
\documentclass[12pt,a4paper]{article}
\usepackage[utf8]{inputenc}
\usepackage[T1]{fontenc}
\usepackage{amsmath,amssymb,amsfonts}
\usepackage{graphicx}
\usepackage{booktabs}
\usepackage{longtable}
\usepackage{pdflscape}
\usepackage{setspace}
\usepackage{array}
\usepackage{multirow}
\usepackage{caption}
\usepackage{subcaption}
\usepackage{placeins}
\usepackage{hyperref}
\usepackage{xurl}
\usepackage{cleveref}
\usepackage{natbib}
\usepackage{xcolor}
\usepackage{listings}
\usepackage{geometry}
\newcommand{\bm}[1]{\boldsymbol{#1}}
\newcommand{\E}{\mathbb{E}}

\DeclareMathOperator*{\argmax}{arg\,max}

\title{Computational Strategies for Mixed Logit Estimation\\[0.5em]
\large Automatic Differentiation, GPU Computing, and Floating-Point Precision
}
\author{
    Hengzhe Zhao\thanks{Macquarie University Centre for the Health Economy (MUCHE) Macquarie University, Sydney, Australia.}\textsuperscript{\;\;}\thanks{Corresponding author. Email: hengzhe.zhao@hdr.mq.edu.au} \and
    Jinhui Zhang\thanks{Macquarie University Business School, Department of Actuarial Studies and Business Analytics, Sydney, Australia.} \and
    Yuanyuan Gu\footnotemark[1]
}

\begin{document}

\maketitle

\begin{abstract}
\setlength{\parskip}{\baselineskip}
\setlength{\parindent}{0pt}
\noindent Mixed logit estimation can become computationally demanding as sample size and the dimension of preference heterogeneity increase. This study examines how derivative computation, processor choice and floating-point precision affect estimation time and parameter recovery across problem sizes. This paper provides a controlled benchmark, comparing finite differences (FD) with automatic differentiation (AD), central processing unit (CPU) execution with graphics processing unit (GPU) execution, and double-precision (FP64) with single-precision (FP32) arithmetic using Monte Carlo experiments that hold the data-generating process, simulation draws, optimiser, starting values, and objective function fixed across strategies. 

The results show that AD is the dominant source of acceleration. Combining AD with FP32 arithmetic provides further gains, while combining AD, GPU execution, and FP32 arithmetic delivers the largest speed improvements, particularly as sample size and model complexity increase. Our findings indicate substantial computational gains with broadly comparable parameter-recovery performance in the simulated settings considered.An empirical application in health preferences illustrates what extra empirical evidence and statistical benefit on model fit could be achieved, when researchers are made possible to estimate a richer, correlated preference specification. 

We provide an open sourced prototype to for model estimation with AD-FP32 strategy workflow that automatically detect hardware and provide the fastest estimation. Faster estimation can make richer random-coefficient specifications, larger simulation draw counts, bootstrap inference, multi-start optimisation, and robustness checks more feasible in applied choice modelling research.
\end{abstract}


\input{sections/01_introduction}

\input{sections/02_computational_background}

\input{sections/03_experimental_design}
\input{sections/04_results}

\input{sections/05_discussion}
\input{sections/06_conclusion}
\clearpage
\appendix
\renewcommand{\thetable}{A\arabic{table}}
\renewcommand{\theHtable}{A\arabic{table}}
\setcounter{table}{0}
\clearpage
\input{sections/appendix_dgp_normalization}
\clearpage
\input{sections/appendix_runtime_tables}
\clearpage
\input{sections/appendix_recovery_tables}
\clearpage
\bibliographystyle{apalike}
\bibliography{references}

\end{document}

%% file: sections/01_introduction.tex
\section{Introduction}

Discrete choice models are widely used to analyse individual preferences and predict choice behaviour across transportation, health, marketing, environmental economics, and many other applied fields \citep{ben-akiva1985discrete,mcfadden1986choice,adamowicz1998stated,lancsar2008conducting}. Among these models, mixed logit has become one of the most widely used approaches because it accommodates flexible substitution patterns and rich preference heterogeneity while remaining behaviourally interpretable \citep{mcfadden2000mixed}. In a recent systematic review of health-related discrete choice experiments published between 2018 and 2023, \citet{nouwens2025evolving} report that 573 of the 1,279 included studies used mixed logit. As empirical applications increasingly seek to estimate willingness to pay (WTP), characterise preference heterogeneity and evaluate complex policy scenarios, mixed logit model has become a standard tool in applied choice modelling.

Applying this model requires researchers to specify how attributes and covariates enter utility, whether their effects are linear or nonlinear, whether higher-order terms or interactions are required, which coefficients vary randomly across individuals, whether random coefficients are correlated, and which mixing distributions are appropriate. These choices can materially affect interpretation: \cite{van2014specification} show that alternative functional forms for utility can change estimated marginal rates of substitution, while \cite{buckell2025break} show that alternative mixing distributions can change estimated preference distributions and WTP. Further considerations include the covariance structure of random coefficients, the number and type of simulation draws, multi-start optimisation, and bootstrap inference, especially in small samples where asymptotic approximations may be less reliable \citep{czajkowski2019simulation,hole2007bootstrap}. Evaluating whether results are credible, interpretable, behaviourally plausible, and robust to reasonable alternatives requires repeated estimation under competing specifications and computational settings, which can become computationally demanding.

Recent work has sought to reduce the computational cost of mixed logit estimation through improvements in both processor choice and gradient computation. Regarding processor choice, \citet{arteaga2022xlogit} demonstrate that graphics processing unit (GPU)-based computation can substantially accelerate mixed logit estimation in xlogit, which uses analytical gradients for optimisation. For gradient computation, finite differences (FD) remain a common fallback when analytical gradients are unavailable in choice-modelling applications \citep{hole2007mixlogit,hess2019apollo}. More recently, \citet{liu2025exploring} show that automatic differentiation (AD) can reduce the computational cost of gradient evaluation relative to FD. Together, these studies demonstrate the potential of GPU acceleration and AD. However, their practical benefits also depend on researchers’ access to hardware and software, differences in hardware capabilities, and how computational methods are implemented. A further consideration is floating-point precision: how precisely numbers are represented and manipulated during computation. These considerations motivate a joint benchmark of computational strategies that can inform both applied estimation and the development of an accessible estimation tool.

These computational dimensions are closely interdependent. AD changes both the computational cost and the numerical properties of gradient evaluation: unlike FD, it avoids repeated likelihood evaluations at perturbed parameter values and the subtraction of nearby log-likelihood values, the latter of which can be particularly sensitive to round-off error under lower-precision arithmetic \citep{higham2002accuracy}. Floating-point precision is closely linked to processor choice. Usually known as single-precision, 32-bit floating-point arithmetic (FP32) requires less memory and can provide substantially higher computational throughput than double-precision, 64-bit floating-point arithmetic (FP64) on many GPU architectures, whereas the relative advantage may be smaller on central processing units (CPUs). Some GPU backends also provide limited or no support for FP64 arithmetic. The computational benefit of GPU execution may therefore depend on floating-point precision, while the numerical consequences of lower precision may depend on how gradients are computed. Yet existing evidence provides limited guidance on how the performance of different combinations varies with sample size and model complexity in mixed logit estimation, or whether their computational gains are accompanied by meaningful changes in parameter recovery.

Motivated by this gap, this study addresses two research questions. First, how do selected combinations of gradient computation (FD versus AD), processor choice (CPU versus GPU), and floating-point precision (FP64 versus FP32) affect the runtime of mixed logit estimation? Second, to what extent do these computational choices affect parameter recovery? To answer these questions, we construct a controlled benchmark with precision-specific stopping tolerances in which the dataset, simulation draws, optimisation algorithm, starting values, and objective function are held fixed within each simulation repetition while the computational configuration is varied. We build on and modify the panel data-generating process (DGP) of \citet{cherchi2012curse}. The original parameterisation is well suited to experiments with a fixed random-coefficient dimension, but extending it to higher dimensions increases the variation in deterministic utility differences and makes choices progressively more deterministic. We therefore modify the original design by rescaling the random coefficients as their number increases, so that the variation in systematic utility differences between alternatives remain approximately constant. The motivation and formal justification for this normalisation are provided in the main text and appendix. This controlled DGP allows us to vary both sample size, \(N\), and model complexity, measured by the number of freely estimated parameters, \(P\), over a broad range of problem sizes.

Because the simulation design has known true parameters, it also allows us to evaluate parameter recovery directly and assess whether faster computational strategies exhibit different recovery performance. This complements previous work such as \citet{liu2025exploring}, which evaluates the numerical accuracy, estimation time, and local-optimum issues associated with alternative gradient calculations. We further complement the simulation benchmark with an empirical stated-preference application in health economics \citep{bertoni2026medical} to examine whether the computational patterns observed under controlled simulation also arise in an applied mixed logit setting. 

Across the experiments, combining AD with GPU execution and FP32 arithmetic delivers the largest runtime reductions under the implemented settings. In one tested setting, this combination achieves a speed-up of over 1,000 relative to FD on a CPU using FP64 arithmetic, reducing mean estimation time from approximately 62 minutes to 3.2 seconds. AD’s advantage over FD increases with model complexity, but substantial gains are also possible without a GPU. In the main benchmark, AD on a CPU outperforms FD on a GPU when both use FP32 arithmetic. Parameter recovery remains broadly similar across the tested strategies, although the largest models still show substantial relative errors in covariance estimates and departures from nominal interval coverage. In the empirical application, AD delivers most of the observed speed-up, while the additional gain from GPU execution is small on the hardware used. Combining AD with GPU execution and FP32 arithmetic allows the fully correlated model to be estimated in approximately 30 seconds. Allowing preference coefficients to correlate changes some WTP estimates and reveals relationships between income sensitivity and job preferences, showing how computational improvements can support richer analyses of choice behaviour.

This study contributes to the choice modelling literature in several ways. First, building on the GPU-based estimation work of \cite{arteaga2022xlogit} and the gradient-method comparisons of \cite{liu2025exploring}, it compares selected combinations of gradient computation, processor choice and floating-point precision within a common estimation framework to assess their conditional and combined performance. Second, it extends the DGP of \cite{cherchi2012curse} to examine how these gains change with sample size and model complexity while keeping the variation in systematic utility differences approximately constant. Unlike a direct expansion of the original parameterisation, this design avoids increasing utility variation simply by adding random coefficients, supporting more interpretable comparisons across model sizes. Third, it complements existing validation exercises in \cite{arteaga2022xlogit} and \cite{liu2025exploring} by jointly assessing estimation error and interval coverage for known mean and covariance parameters across repeated simulated datasets under different computational configurations. Finally, it extends the empirical analysis of \cite{bertoni2026medical} to examine the behavioural implications of a fully correlated preference specification. A software prototype requiring minimal coding supports the practical application of these methods and can be used locally or through an online interface (https://huggingface.co/spaces/Wil2200/prefero).

The rest of the paper is organised as follows. Section 2 introduces the computational background, covering mixed logit estimation, FD and AD gradient computation, floating-point precision and processor choice. Section 3 introduces the benchmark design and DGP. Section 4 presents the runtime and parameter recovery results, together with a case study illustrating how a fully correlated mixed logit model can provide additional insights into an empirical application. Section 5 discusses the implications and limitations of the research, and Section 6 concludes.

%% file: sections/02_computational_background.tex
\section{Computational Background}
\label{sec:methodology}

This section introduces the computational aspects underlying the benchmark. We first describe the mixed logit model and the simulated log-likelihood function used for estimation. We then consider the three dimensions examined in this study: gradient computation, floating-point precision, and processor choice. These dimensions affect runtime through different but interrelated mechanisms.

\subsection{Mixed logit estimation}

Mixed logit models can be used to analyse both revealed and stated preference data. Consider individual $n \in 1,\ldots,N$  observed over $T$ choice occasions, each with $J$ alternatives. The utility that individual $n$ assigns to alternative $j$ on occasion $t$ is

\begin{equation}
  U_{ntj} = \bm{x}_{ntj}'\bm{\beta}_n + \varepsilon_{ntj},
\qquad
\varepsilon_{ntj} \overset{i.i.d.}{\sim} \mathrm{EV1},
  \label{eq:utility}
\end{equation}

where $\bm{x}_{ntj}$ is the vector of attributes that describe the alternative ($j$) in occasion ($t$) and $\bm{\beta}_n$ is the individual-specific vector for preference coefficients. The errors follow an independent and identically distributed type-I extreme-value distribution. Conditional on $\bm{\beta}_n$ and under the assumed error distribution, the probability of choosing alternative $j$ in occasion $t$ is the multinomial logit probability

\begin{equation}
  P_{ntj}(\bm{\beta}_n) = \frac{\exp(\bm{x}_{ntj}'\bm{\beta}_n)}{\sum_{k=1}^{J} \exp(\bm{x}_{ntk}'\bm{\beta}_n)}.
  \label{eq:logit_kernel}
\end{equation}

Mixed logit treats $\bm{\beta}_n$ as a draw from a population distribution. This allows preferences to vary across individuals while retaining a common structure for that heterogeneity. The likelihood contribution of individual $n$ for the observed sequence $\bm{y}_n=(y_{n1},\ldots,y_{nT})$ is obtained by integrating the product of conditional logit probabilities over the mixing distribution,

\begin{equation}
  P_n(\bm{\theta}) = \int \prod_{t=1}^{T} P_{nt,y_{nt}}(\bm{\beta})\, f(\bm{\beta} \mid \bm{\theta})\, d\bm{\beta}.
  \label{eq:panel_probability}
\end{equation}

We illustrate the estimation procedure using a multivariate Gaussian mixing distribution, a common choice in empirical mixed logit applications. Let $\bm{\theta}=(\bm{\mu}',\operatorname{vech}(\bm{L})')'$ collect the mean vector $\bm{\mu}$ and Cholesky parameters $\bm{L}$ which defines the covariance matrix (\(\omega = LL'\)). Conditional on a normal draw $\bm{z}_n$, the coefficient vector implied by $\bm{\theta}$ can be written as

\[
  \bm{\beta}_n(\bm{\theta}) = \bm{\mu} + \bm{L}\bm{z}_n,
  \qquad
  \bm{z}_n \sim \mathcal{N}(\bm{0},\bm{I}).
\]

The integral in \Cref{eq:panel_probability} generally has no closed-form expression, so it is approximated with Monte Carlo simulation. For simulation draw $r$ of individual $n$, 
\[
  \bm{\beta}_n^{(r)}(\bm{\theta}) = \bm{\mu} + \bm{L}\bm{z}_n^{(r)},
  \qquad
  \bm{z}_n^{(r)}\sim\mathcal{N}(\bm{0},\bm{I}).
\]

Averaging over $R$ simulation draws gives the simulated probability

\begin{equation}
  \widehat{P}_n(\bm{\theta}) = \frac{1}{R}\sum_{r=1}^{R} \prod_{t=1}^{T} P_{nt,y_{nt}}\!\left(\bm{\beta}_n^{(r)}(\bm{\theta})\right),
  \label{eq:simulated_probability}
\end{equation}

and summing its logarithm across individuals gives the simulated log-likelihood,

\begin{equation}
  \ell^{S}(\bm{\theta}) = \sum_{n=1}^{N} \log \widehat{P}_n(\bm{\theta}).
  \label{eq:sll}
\end{equation}

\Cref{eq:sll} is the objective function maximised throughout the benchmark, and its repeated evaluation is the main source of computational cost. The cost of a single likelihood evaluation increases with the number of individuals, choice occasions, alternatives, simulation draws, and random coefficients. Total runtime additionally depends on the number of likelihood evaluations required during optimisation. This is particularly important under finite differences, for which the number of objective-function evaluations required to construct a gradient increases directly with the number of freely estimated parameters.

Runtime can be reduced through two main channels: reducing the number of likelihood evaluations required to obtain a gradient and reducing the time required for each likelihood evaluation. The first channel is primarily affected by gradient computation, whereas the second is influenced by processor choice and floating-point precision. These two channels motivate the scaling experiments in \Cref{sec:experimental_design}, in which we vary both sample size and model complexity.

\subsection{Gradient computation}
The simulated maximum-likelihood estimator is
\[\widehat{\bm{\theta}} = \argmax_{\bm{\theta}} \ell^{S}(\bm{\theta}).\]
During optimisation, the optimiser takes the gradient of \(\ell^{S}\) with respect to \(\bm{\theta}\) to update the parameter vector. The method used to compute this gradient can therefore have a substantial effect on the runtime. Under FD, the gradient with respect to parameter $\theta_i$ is approximated as
\begin{equation}
\frac{\partial \ell^{S}}{\partial \theta_i}
\approx
\frac{
\ell^{S}(\bm{\theta} + h\,\bm{e}_i)
-
\ell^{S}(\bm{\theta} - h\,\bm{e}_i)
}
{2h},
\label{eq:fd}
\end{equation}

where \(\bm{e}_i\) is the \(i\)th basis vector and \(h\) is the FD step size. FD is broadly applicable because it only requires evaluations of \(\ell^{S}\), but generality comes at substantial computational cost. For a model with \(P\) freely estimated parameters, FD scheme requires approximately \(2P\) evaluations of \(\ell^{S}\) to construct one full gradient. Its computational cost is therefore massive and increases directly with model complexity, which is especially important for mixed logit models with many random coefficients which introduce freely estimated covariance parameters.

The choice of \(h\) also creates a numerical trade-off. If \(h\) is too large, the FD approximation can have substantial truncation error. If \(h\) is too small, the calculation subtracts two increasingly similar log-likelihood values, increasing the influence of round-off error. This cancellation can make FD less reliable in lower-precision arithmetic and limit the computational gains available from using fewer significant digits \citep{higham2002accuracy}.

Rather than perturbing individual parameters and repeatedly re-evaluating the objective function, AD scheme records the sequence of elementary operations used to compute \(\ell^{S}\), and apply chain rule backwards through this sequence.  It therefore computes gradient of the implemented likelihood directly, provided that the operations involved are differentiable. The distinction between FD and AD is illustrated in \Cref{fig:fd_ad_example}.

For mixed logit estimation, AD is particularly well suited because it calculates the derivatives of a single log-likelihood function with respect to potentially many parameters. It computes the gradient without evaluating the log-likelihood twice for each parameter, as required by FD. Analytical gradients can similarly avoid these repeated evaluations, but they must be derived and implemented for each model specification hence is not ideal for package implementation. AD instead differentiates the implemented likelihood automatically, reducing the need to derive and code model-specific gradients manually.

AD also removes the FD's step-size choice and avoids the subtraction of nearby likelihood values. These properties link gradient computation directly to floating-point precision: the computational advantage of AD can become more pronounced as \(P\) increases, while its numerical properties may also make lower-precision arithmetic more viable than under FD.

\begin{center}
\begin{minipage}{0.96\textwidth}
\centering
\fbox{%
\begin{minipage}{0.94\textwidth}
\small
\textbf{Example.} Compute the derivative of $f(x)=6x^2$ at $x=2$.

\vspace{0.75em}
\begin{minipage}[t]{0.46\textwidth}
\textbf{Finite Differences (FD)}

Choose $h=0.001$ and evaluate the function at two nearby points:
\[
\begin{aligned}
f(2+h) &= 24.024006,\\
f(2-h) &= 23.976006.
\end{aligned}
\]
The derivative is approximated by the slope between these two evaluations:
\[
\frac{df}{dx}
\approx
\frac{f(2+h)-f(2-h)}{2h}
=24.
\]
\end{minipage}
\hfill
\begin{minipage}[t]{0.46\textwidth}
\textbf{Automatic Differentiation (AD)}

Write the same function as elementary operations:
\[
u=x^2,\qquad f=6u.
\]
AD propagates derivatives through these operations using the chain rule:
\[
\begin{aligned}
\frac{du}{dx} &= 2x,\\
\frac{df}{du} &= 6,\\
\frac{df}{dx}
&=
\frac{df}{du}\cdot\frac{du}{dx}
=6(2x)=24.
\end{aligned}
\]
\end{minipage}
\end{minipage}
}
\captionsetup{type=figure,hypcap=false}
\caption{Derivative computation by FD and AD in an example.}
\label{fig:fd_ad_example}
\end{minipage}
\end{center}

\subsection{Floating-point precision}

Under FD, each gradient is constructed from differences between log-likelihood values obtained by slightly increasing and decreasing each parameter. Floating-point precision determines how numbers are stored during computation. In our work, we consider two most popular precisions FP64 and FP32. FP64 retains approximately 16 decimal digits of precision, whereas FP32 retains approximately 7 \citep{higham2002accuracy}. 

FP32 also requires less memory per value and utilise the hardware to compute more efficiently. These computational advantages come at the cost of reduced numerical precision and greater sensitivity to round-off error. Lower precision can therefore make the subtraction in \Cref{eq:fd} less reliable, particularly when difference between these values is small relative to their magnitudes. AD does not rely on the subtraction of nearby log-likelihood values and may therefore be better suited to lower-precision computation. AD nevertheless remains subject to floating-point round-off error.

Precision is also closely linked to processor choice. The relative throughput of FP32 and FP64 is hardware dependent: CPUs and GPUs differ in how strongly they are optimised for each format, and some accelerator backends provide limited or no support for FP64 arithmetic. Floating-point precision is therefore not only a numerical setting but also a computational choice that can influence which processor configurations are available and how effectively their computational capacity can be used. For this reason, the benchmark evaluates floating-point precision jointly with gradient computation and processor choice. AD nevertheless remains subject to floating-point round-off error.

\subsection{Processor choice}

As the third computational dimension, we considered processor choice of CPU and GPU in the benchmark. CPUs are designed for general-purpose computation and remain the conventional platform for most econometric estimation. GPUs, by contrast, contain many processing units designed to execute large numbers of similar operations in parallel. This architecture is well suited to numerical workloads containing repeated and largely independent calculations \citep{kirk2016programming}.

Simulation-based mixed logit estimation has several characteristics that make such parallelism potentially useful. Likelihood evaluation repeats similar calculations across respondents, choice tasks, alternatives, and simulation draws. Once the data, parameters, and simulation draws are fixed for a likelihood evaluation, many of these calculations can be performed simultaneously. The potential benefit of GPU execution can also increase as the computational workload grows with sample size, the number of simulation draws, or model complexity.

However computing with GPU introduces extra runtime to initialise, including kernel launches, memory allocation, and, theoretically depending on the support software implementation, data transfer between CPU and GPU memory. For sufficiently small estimation problems, these initial cost can offset part of the reduce in runtime from switching to GPU. However as what is shown later in result, GPU perform faster even on small sample size and models in the field of mixed logit estimation. 

Processor choice also interacts with floating-point precision. Many GPUs provide substantially greater power for FP32 than for FP64, while some GPU software backends provide limited or no support for FP64 arithmetic. Consequently, the value of GPU execution depends partly on whether mixed logit estimation can be performed reliably using FP32. Overall, this completes the link among the three computational dimensions examined in the benchmark: gradient computation affects the numerical suitability of lower precision, floating-point precision affects the computational advantage and availability of GPU execution, and processor choice determines how effectively likelihood calculations can exploit parallel computation.

%% file: sections/03_experimental_design.tex
\section{Experimental Design}
\label{sec:experimental_design}

The benchmark combines three computational dimensions: gradient computation using FD or AD, processor execution on a CPU or GPU, and floating-point arithmetic using FP64 or FP32. FD-CPU-FP64 is used as the reference strategy because it represents a conventional configuration based on numerical gradients, CPU execution, and FP64. Because many commodity GPUs devote substantially greater computational capacity to FP32 than to FP64 \citep{jia2018dissecting}, GPU-FP64 strategies are not included in the benchmark. The benchmark focuses on GPU-FP32 execution, while retaining both FP64 and FP32 on the CPU to isolate the effect of precision when GPU acceleration is unavailable. 

The optimisation algorithm is held fixed across strategies. All models are estimated using PyTorch's L-BFGS optimiser with a strong Wolfe line search, a history size of 100, a maximum of 500 iterations, and the same zero starting vector. We use a looser gradient stopping tolerance for FP32 (\(10^{-2}\)) than for FP64 (\(10^{-5}\)) since FP32 arithmetic retains approximately seven significant decimal digits, compared with approximately sixteen for FP64. We believe that setting lower tolerance for FP32 is trivial for convergence because FP32 itself limits the numerical resolution of likelihood and gradient calculations become noisy when near convergence. Because the simulated datasets have true parameter values from DGP, we can assess estimation accuracy directly through parameter recovery, reporting mean absolute error (MAE), mean absolute percentage error (MAPE) and interval coverage across computational strategies.

The simulation experiments use 50 Monte Carlo repetitions. Each repetition begins with a single random seed that is used throughout, including the simulated panel, respondent-level random coefficients, observed choices, and Modified Latin Hypercube Sampling (MLHS) draws used to approximate the likelihood. Throughout the simulation benchmark, the likelihood is approximated using 1,000 MLHS draws. MLHS draw method was used because it efficiently reduce simulation errors for given number of draws \citep{hess2006mlhs,cherchi2012curse}. We used zero starting values throughout.

All experiments are conducted on an Amazon Web Services EC2 G6.4xlarge instance with 16 vCPUs, 64 GiB of RAM, and one NVIDIA L4 GPU with 22.3 GiB of GPU memory as detected by PyTorch. The software environment consists of Windows Server 2022, Python 3.13.11, and PyTorch 2.6.0+cu124 with CUDA 12.4. For each run, we record the runtime to convergence, followed by evaluating parameter recovery by comparing the estimated parameters with true parameter values. The DGP is described in \Cref{sec:simulated_panel}, and the recovery measures are defined in \Cref{sec:parameter_recovery_metrics}.

\subsection{Main benchmark and scaling experiments}

The main benchmark uses a simulated dataset with \(N=1{,}000\) respondents. Each respondent faces \(T=8\) choice tasks, each task contains \(J=3\) alternatives, and each alternative is described by five attributes. We estimate a mixed logit model with five normally distributed random coefficients and a full covariance structure parameterised through its Cholesky factor. The model consequently contains five mean parameters and 15 Cholesky parameters, giving \(P=20\) freely estimated parameters in total. Alternative-specific constants are omitted from both the DGP and the estimated model. This benchmark evaluates all six computational strategies under a common stated-preference setting and provides the baseline comparison of gradient computation, processor choice, and floating-point precision.

We then conduct two scaling experiments to examine how the relative performance of these strategies changes as the estimation problem becomes larger regarding two different measures. The first varies sample size while holding model complexity fixed, whereas the second varies model complexity while holding sample size fixed. Despite the implication for testing generalisability for empirical problems at different sizes, computationally, this distinction is also important because increasing \(N\) primarily enlarges the amount of likelihood computation performed for a fixed parameter vector, while increasing \(P\) also changes the cost of gradient computation and the dimensionality of the problem.

The \(N\)-scaling experiment varies the number of respondents over $N \in$ \{200, 500, 1,000, 2,000, 5,000, 10,000\} while holding the model specification fixed at five random coefficients and \(P=20\) freely estimated parameters. The \(P\)-scaling experiment instead holds the sample size fixed at \(N=1{,}000\) and increases the dimension of the random-coefficient specification. Let \(p_{\mathrm{mean}}\) denote the number of random-coefficient means and let $p_{\mathrm{vcov}}$ denote the number of free Cholesky parameters under the full covariance specification. The total number of freely estimated parameters \(P\) becomes

$$
P=p_{\mathrm{mean}}+ p_{\mathrm{vcov}},  
\quad 
\text{where } p_{\mathrm{vcov}} = \frac{p_{\mathrm{mean}}(p_{\mathrm{mean}}+1)}{2}
$$

The experiment considers \(P\in\{20,65,135,230\}\), corresponding respectively to \(p_{\mathrm{mean}}\in\{5,10,15,20\}\) random coefficients, or equivalently, number of attributes. The baseline case contains five mean parameters and 15 Cholesky parameters, while the largest specification contains 20 mean parameters and 210 Cholesky parameters.

Unlike the main benchmark, the scaling experiments only compare AD-CPU-FP64, AD-CPU-FP32, and AD-GPU-FP32 against FD-CPU-FP64 as the reference strategy. FD-CPU-FP32 and FD-GPU-FP32 are excluded because both the findings from our N-scaling experiment showed these two remain inferior to their AD counterpart, which is consistent with literature indicate that AD outperforms FD \citep{liu2025exploring}. Including these two FD strategies would also consume substantial computational resources: as shown later, FD-based estimation is extremely time-consuming for models with complex correlation structures. We retain FD-CPU-FP64 as a baseline to represent software configurations commonly available to researchers.

\subsection{Data-generating process}
\label{sec:simulated_panel}

Our simulation is done with a DGP, reivsed from the one used in \citet{cherchi2012curse}. In our DGP, each simulated individual observed \(T=8\) choice occasions, each with \(J=3\) alternatives. For individual \(n\), choice occasion \(t\), alternative \(j\), and attribute dimension \(m\), attributes are drawn independently as
\begin{equation*}
    x_{ntjm} \sim \mathcal{TN}_{[0,3]}(1.5,0.9^2),
    \label{eq:generic_attribute_dgp}
\end{equation*}
where \(\mathcal{TN}_{[0,3]}(1.5,0.9^2)\) denotes a normal distribution with location \(1.5\) and scale \(0.9\), truncated to the interval \([0,3]\). The bound keeps the simulated attributes in a realistic range while preserving random variation across alternatives.

Individual-level coefficients are generated as
\begin{equation*}
    \bm{\beta}_{n} = \bm{\mu}_0 + \bm{L}_0\bm{z}_{n},
    \qquad
    \bm{z}_{n} \sim \mathcal{N}(\bm{0},\bm{I}),
    \label{eq:random_coefficients_dgp}
\end{equation*}
where \(\bm{\mu}_0\) is the true mean vector and \(\bm{L}_0\) is the lower-triangular Cholesky factor of the true covariance matrix, so that
\[
    \bm{\Omega}_0 = \bm{L}_0\bm{L}_0'.
\]

The original parameterisation assigns a common mean of (-1) to each random coefficient and uses a covariance matrix with diagonal elements (0.3) and off-diagonal elements (0.1). This parameterisation is suitable when the random-coefficient dimension is fixed, as in the (N)-scaling experiment. However, extending it directly to the (P)-scaling experiment, would change the underlying choice problem as model complexity increases. Holding the magnitude of each coefficient fixed while adding random coefficients increases the variance of deterministic utility differences. Because the scale of the Gumbel error remains unchanged, choices become progressively more deterministic as the random-coefficient dimension increases, and the resulting likelihood can become easier to optimise. Runtime differences across (P)-scaling cells would then reflect both greater model complexity and a simultaneous change in the underlying choice difficulty.

To mitigate these effects, we normalise the magnitude of the random coefficients as dimension increases. For alternatives (j) and (k) in task (t), let \(\Delta V_{ntjk}=V_{ntj}-V_{ntk}\) denote their deterministic utility difference. We use \(\operatorname{Var}(\Delta V)\) to denote the variance of these pairwise utility differences over simulated individuals, choice tasks, and alternative pairs. Under the indenpendency assumption with attribute design, \(\operatorname{Var}(\Delta V)\) is proportional to the second moment of the random-coefficient vector, as shown in Appendix \ref{app:dgp_normalization}. We normalise the true coefficient distribution so that

\begin{equation}
  \E[\|\bm{\beta}_n\|^2] = \|\bm{\mu}_0\|^2+\operatorname{tr}(\bm{L}_0\bm{L}_0')=1.
\end{equation}

A share \(1-\lambda\) of this second moment is assigned to the squared mean and a share \(\lambda\) to heterogeneity,
\begin{equation*}
    \|\bm{\mu}_0\|^2 = 1-\lambda,
    \qquad
    \operatorname{tr}\left(\bm{L}_0\bm{L}_0'\right)=\lambda,
    \qquad
    \lambda=\tfrac{1}{4}.
\end{equation*}

Let \(\widetilde{\bm{\mu}}\) and \(\widetilde{\bm{L}}\) denote the unnormalised mean vector and Cholesky factor. Specifically, \(\widetilde{\bm{\mu}}=-\bm{1}\), and \(\widetilde{\bm{L}}=\operatorname{chol}(\widetilde{\bm{\Omega}})\), where \(\widetilde{\bm{\Omega}}\) has diagonal elements \(0.30\) and off-diagonal elements \(0.10\). The normalised true values used in our DGP are
\begin{equation*}
    \bm{\mu}_0
    =
    \sqrt{\frac{1-\lambda}{\|\widetilde{\bm{\mu}}\|^2}}
    \,\widetilde{\bm{\mu}}
    =
    -\sqrt{\frac{1-\lambda}{p_{\mathrm{mean}}}}\bm{1},
    \qquad
    \bm{L}_0
    =
    \sqrt{
    \frac{\lambda}
    {\operatorname{tr}(\widetilde{\bm{L}}\widetilde{\bm{L}}')}
    }
    \,\widetilde{\bm{L}}.
\end{equation*}

The magnitudes of the true mean and Cholesky parameters decrease as the number of random coefficients increases, while the total second moment of $\bm{\beta}_n$ remains fixed. \Cref{tab:dgp_difficulty} shows the resulting variation in pairwise deterministic utility differences in our simulated dataset. Under the normalised DGP, $\operatorname{Var}(\Delta V)$ remains approximately constant as (P) increases. Observed choices are generated by adding independent standard Gumbel errors to the deterministic utilities and selecting the alternative with the highest total utility.

\begin{table}[tbp]
\centering
\caption{Pairwise deterministic utility variation by model complexity.}
\label{tab:dgp_difficulty}
\small
\begin{tabular}{rcc}
\toprule
 & \multicolumn{2}{c}{\(\operatorname{Var}(\Delta V)\)} \\
\cmidrule(lr){2-3}
\(p\) & Normalised DGP & Unnormalised DGP \\
\midrule
20  & 1.03 & 6.69 \\
65  & 1.02 & 13.29 \\
135 & 1.03 & 20.13 \\
230 & 1.02 & 26.56 \\
\bottomrule
\end{tabular}
\medskip
\end{table}

\subsection{Parameter recovery metrics}
\label{sec:parameter_recovery_metrics}

In addition to runtime, we evaluate whether each computational strategy recovers the parameters of the DGP. For each strategy in a given experiment and each Monte Carlo repetition, let \(\widehat{\bm{\mu}}\) and \(\widehat{\bm{L}}\) denote the estimated mean vector and Cholesky factor, and define \(\widehat{\bm{\Omega}}=\widehat{\bm{L}}\widehat{\bm{L}}'\). We assess recovery of the true mean vector \(\bm{\mu}_0\) using \(\widehat{\bm{\mu}}\), and recovery of the true covariance matrix \(\bm{\Omega}_0\) using \(\widehat{\bm{\Omega}}\).

For the variance--covariance structure, recovery is evaluated using \(\widehat{\bm{\Omega}}\) rather than \(\widehat{\bm{L}}\), even though the optimiser estimates Cholesky parameters. In the simulation design, specifying the true covariance matrix \(\bm{\Omega}_0\) or its Cholesky factor \(\bm{L}_0\) is equivalent because entries of both matrix are semidefinite. In estimation, however, entries of \(\widehat{\bm{L}}\) could be negative during optimisation, while representing same covariance matrix. We consequently evaluate covariance recovery in \(\bm{\Omega}\)-space.

For the mean parameters, we report the mean absolute error (MAE) and mean absolute percentage error (MAPE) across the \(p_{\mathrm{mean}}\) random-coefficient means,
\begin{align}
    \operatorname{MAE}_{\mu}
    &=
    \frac{1}{p_{\mathrm{mean}}}\sum_{m=1}^{p_{\mathrm{mean}}}
    \left|\widehat{\mu}_{m}-\mu_{0m}\right|,
    \\
    \operatorname{MAPE}_{\mu}
    &=
    \frac{100}{p_{\mathrm{mean}}}\sum_{m=1}^{p_{\mathrm{mean}}}
    \frac{\left|\widehat{\mu}_{m}-\mu_{0m}\right|}{\left|\mu_{0m}\right|}.
\end{align}
For the covariance matrix, we compute the same measures across the unique variance--covariance elements,
\begin{align}
    \operatorname{MAE}_{\Omega}
    &=
    \frac{1}{p_{\mathrm{vcov}}}\sum_{k=1}^{p_{\mathrm{vcov}}}
    \left|
    \operatorname{vech}(\widehat{\bm{\Omega}})_k
    -
    \operatorname{vech}(\bm{\Omega}_{0})_k
    \right|,
    \\
    \operatorname{MAPE}_{\Omega}
    &=
    \frac{100}{p_{\mathrm{vcov}}}\sum_{k=1}^{p_{\mathrm{vcov}}}
    \frac{\left|
    \operatorname{vech}(\widehat{\bm{\Omega}})_k
    -
    \operatorname{vech}(\bm{\Omega}_{0})_k
    \right|}{\left|\operatorname{vech}(\bm{\Omega}_{0})_k\right|}.
\end{align}

Both MAE and MAPE are computed within each Monte Carlo repetition and then averaged across the 50 repetitions, with sample standard deviations reported in parentheses. These measures summarise point-estimation error but do not account for finite-sample uncertainty, which requires evaluating interval coverage. For each estimated parameter, we construct an approximate 95\% confidence interval using Berndt--Hall--Hall--Hausman (BHHH) standard errors \citep{berndt1974bhhh} and record whether the corresponding true value lies within that interval. For covariance parameters, standard errors for the estimated Cholesky elements are transformed into standard errors for \(\operatorname{vech}(\widehat{\bm{\Omega}})\) using the delta method. Coverage is then reported as the proportion of true parameter values contained within the corresponding 95\% intervals.

%% file: sections/04_results.tex
\section{Results}
\label{sec:results}

This section reports the runtime and parameter recovery results from the benchmark experiments. The main text focuses on the main benchmark runtime comparison, the two runtime scaling patterns, and a summary of parameter recovery. Full runtime tables for scaling experiments are reported in Appendix \ref{app:runtime_tables}, while full recovery statistics are reported in Appendix \ref{app:recovery_tables}.

\subsection{Main benchmark}

\label{sec:results:main_benchmark}

\Cref{tab:baseline_runtime} reports runtime and speed-up for the main benchmark with \(N=1{,}000\) respondents and \(P=20\) freely estimated parameters. The reference strategy, FD-CPU-FP64, has a mean runtime of 331.3 seconds (variance of 101.8 seconds) across the 50 Monte Carlo repetitions. Replacing FD with AD substantially reduces runtime without changing processor choice or floating-point precision: AD-CPU-FP64 has a mean runtime of 45.0 seconds (variance of 13.5 seconds), corresponding to a \(7.35\times\) speed-up.

Further gains arise when AD is combined with FP32 arithmetic. AD-CPU-FP32 reduces the mean runtime to 7.7 seconds (variance of 2 seconds), corresponding to a \(42.57\times\) speed-up relative to FD-CPU-FP64. Among the six strategies considered, AD-GPU-FP32 has the lowest mean runtime at 1.5 seconds (variance of 0.4 second) and achieves a \(219.42\times\) speed-up relative to the reference strategy.

The FD-based strategies also benefit from FP32 arithmetic and GPU execution, although their gains are smaller than those achieved by the corresponding AD-based strategies. FD-CPU-FP32 reduces the mean runtime to 144.8 seconds (variance of 41.2 seconds), while FD-GPU-FP32 reduces it to 22.8 seconds (variance of 7.3 seconds). FD-GPU-FP32 nevertheless remains substantially slower than AD-GPU-FP32, indicating that GPU execution does not eliminate the additional computational cost associated with FD gradient evaluation. Overall, the main benchmark shows that replacing FD with AD provides the largest reduction in runtime, while FP32 arithmetic and GPU execution provide further gains when combined with AD. Meanwhile non-trivial differences between strategies were not found in any of the MAE, MAPE, and interval coverage as listed in Appendix Table \ref{tab:app_baseline_recovery}. This has shown that the runtime reduce is achieved without materially different estimation results, even with stopping gradient being set differently.

\begin{table}[htbp]
\centering
\caption{Main benchmark runtime and speed-up
(\(N=1{,}000\), \(P=20\); 50 Monte Carlo repetitions).}
\label{tab:baseline_runtime}
\small
\begin{tabular}{lrr}
\toprule
Strategy & Runtime (s) & Speed-up \\
\midrule
FD-CPU-FP64 (ref.) & 331.3 (101.8) & 1.00 (--) \\
AD-CPU-FP64        & 45.0 (13.5)   & 7.35 (0.51) \\
AD-CPU-FP32        & 7.7 (2.0)     & 42.57 (12.77) \\
AD-GPU-FP32        & 1.5 (0.4)     & 219.42 (67.02) \\
FD-CPU-FP32        & 144.8 (41.2)  & 2.27 (0.86) \\
FD-GPU-FP32        & 22.8 (7.3)    & 14.55 (5.62) \\
\bottomrule
\end{tabular}
\medskip

\footnotesize\textit{Note:}
Runtimes are means across the 50 Monte Carlo repetitions, with sample standard
deviations in parentheses. Speed-ups are geometric means of within-repetition
runtime ratios relative to FD-CPU-FP64, with sample standard deviations of those
ratios in parentheses.

\end{table}

\subsection{Scaling experiments}
\label{sec:results:n_scaling}

When we vary the number of individuals \(N\) from 200 to 10,000 while holding others constant, the mean runtime increases with sample size for all four strategies, while the runtime gap between FD-CPU-FP64 and the AD-based strategies becomes larger as \(N\) increases as shown in \Cref{fig:n_scaling_runtime} of which the numbers are listed in appendix table \ref{tab:app_n_scaling_runtime}. At \(N=200\), FD-CPU-FP64 requires 58.4 seconds on average, compared with 0.7 seconds for AD-GPU-FP32, corresponding to an \(88.62\times\) speed-up. At \(N=10{,}000\), the corresponding mean runtimes are 3,370.9 seconds and 6.9 seconds, respectively, with AD-GPU-FP32 achieving a \(493.75\times\) speed-up relative to FD-CPU-FP64.

The relative advantage of AD-GPU-FP32 increases with sample size. This pattern is consistent with the parallel structure of the simulated likelihood described in Section~\ref{sec:methodology}: as \(N\) increases, each likelihood evaluation involves a larger number of indvidual-level calculations that can be computed in parallel on the GPU. Substantial runtime reductions are also observed without use of GPU. Across the six sample sizes considered, AD-CPU-FP32 achieves speed-ups ranging from \(32.32\times\) to \(42.99\times\) relative to FD-CPU-FP64.

\begin{figure}[!htb]
\centering
\includegraphics[width=0.82\textwidth]{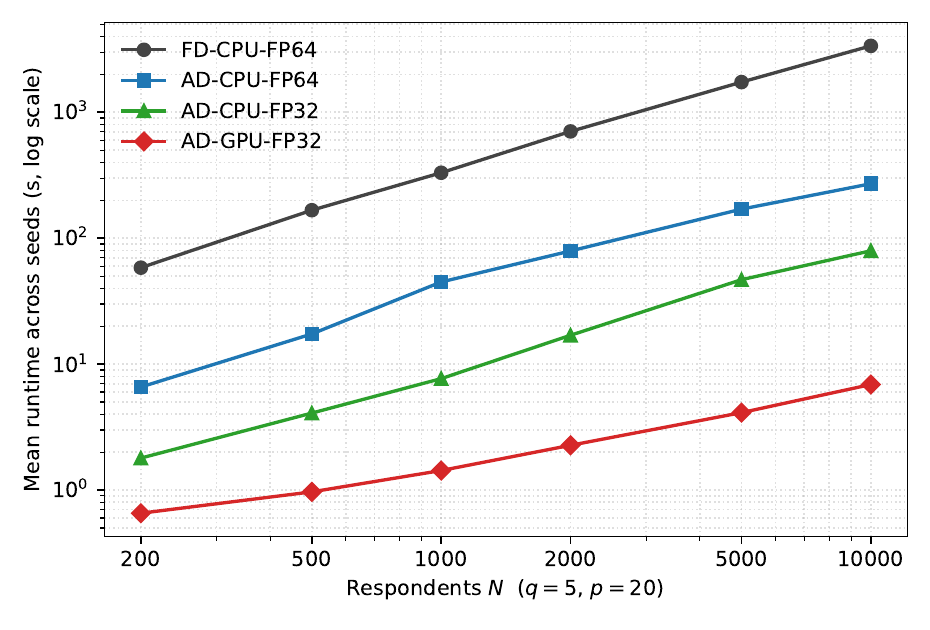}
\caption{Runtime by sample size in the \(N\)-scaling experiment. Runtime is reported in seconds on a logarithmic scale. Each point is the mean across 50 Monte Carlo repetitions; \(P=20\) for all models.}
\label{fig:n_scaling_runtime}
\end{figure}

Alternativly when we vary the model complexity by adjusting the number of freely estimated parameters from \(P=20\) to \(P=230\), while fixing number of individuals \(N\) at 1,000. The runtime differences between FD-CPU-FP64 and the AD-based strategies increase as model complexity increases as shown in \Cref{fig:p_scaling_runtime} of which the numbers are listed in appendix table \ref{tab:app_p_scaling_runtime}. At \(P=20\), FD-CPU-FP64 has a mean runtime of 331.2 seconds. At \(P=65\), its mean runtime increases to 3,718.9 seconds. Over the same increase in model complexity, AD-CPU-FP64 increases from 45.0 to 142.7 seconds, AD-CPU-FP32 from 7.7 to 20.3 seconds, and AD-GPU-FP32 from 1.5 to 3.2 seconds. For AD-GPU-FP32, the speed-up increases from \(219.28\times\) at \(P=20\) to \(1{,}147.4\times\) at \(P=65\). For AD-CPU-FP32, it increases from \(42.61\times\) to \(183.17\times\) over the same range. FD-CPU-FP64 was not estimated for \(P=135\) and \(P=230\) because of the high computational cost, which is still minor for other AD-based strategies as seen in the figure.

This pattern is consistent with the difference in gradient-computation cost described in Section~\ref{sec:methodology}. Under FD, constructing a full gradient requires \(2P\)objective-function evaluations, whereas AD does not require separate parameter-by-parameter perturbations. The runtime gap between FD-CPU-FP64 and the AD-based strategies therefore becomes larger as the number of freely estimated parameters increases.

Lastly, for every N or P, all strategies tested only showed arbitrarily small difference in recovering mean and covaraince parameters, across three metrics used as shown in table~\ref{tab:app_n_scaling_recovery} and table~\ref{tab:app_p_scaling_recovery}. As N goes larger, reduce in MAE and MAPE is observed subject to asymptotic properties of our estimator \citep{mcfadden2000mixed}. At the same time, coverage rate reduced a little but remained considerablly high over 93\% overall. As sample size increase, standard error and confidence interval becomes smaller in absolute scale from increasing amount of information, however it leads to higher statistical certainty such that a narrower interval is less likely to capture the true value. In empirical practice, this issue can be addressed by using multi-starting values, which is regardless, not the focus of our research. For P-scaling experiment, MAE and MAPE goes higher when P goes higher subject to realtivly less information exist when sample size stay same while model complexity grows. Especially for MAPE, due to normalization in DGP, true value of covariance parameter becomes smaller and thus the denominators in MAPE, making it more sensitive to extreme estimated values. Nevertheless, coverage rate remained satisfically high, and these issues are mitigated with multi-start estimations of which the high computational cost constraints can be mitigated by applying a sound strategy, informed by our study.

\begin{figure}[!htb]
\centering
\includegraphics[width=0.78\textwidth]{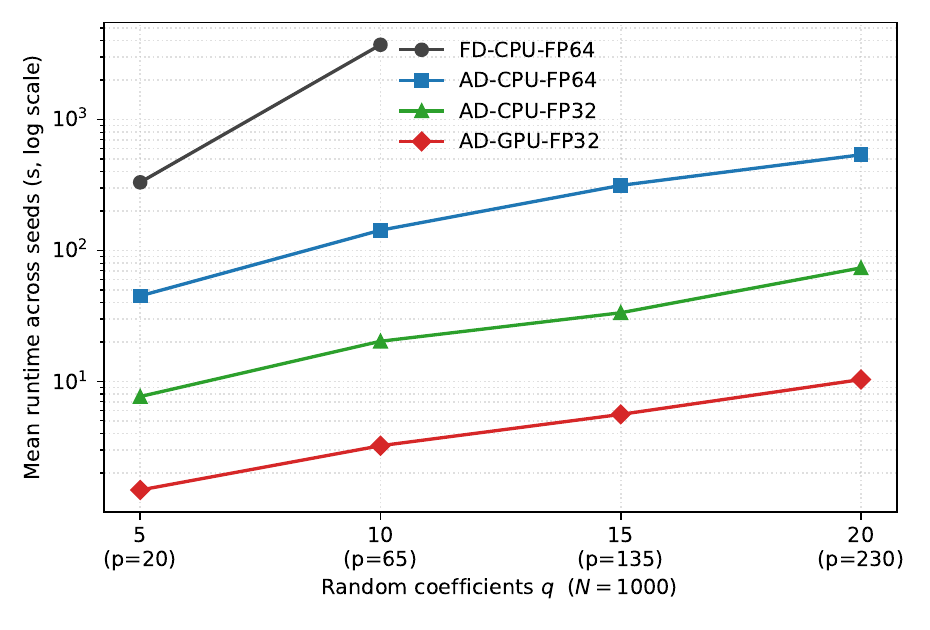}
\caption{Runtime by model complexity in the \(P\)-scaling experiment. Runtime is reported in seconds on a logarithmic scale. Each point is the mean across 50 Monte Carlo repetitions; \(N=1{,}000\) for all models. FD-CPU-FP64 was not estimated for \(P=135\) and \(P=230\) because of high computational cost.}
\label{fig:p_scaling_runtime}
\end{figure}

\subsection{Case study: Italian medical students' job choices}
\label{sec:medical_application}

We apply the computational strategies to Italian medical students' stated job choices \citep{bertoni2026medical} for three purposes: to assess whether the computational gains carry over to observed data, to examine performance with relatively long respondent panels, and to estimate a richer preference specification that was previously computationally costly to fit.\footnote{We sincerely thank participants at the 2026 International Choice Modelling Conference for suggesting an empirical application with longer choice-task sequences. This dataset contains 16 tasks per respondent and provides a useful observed-data test of the computational methods developed in this paper and its implication for empirical analysis.} Medical brain drain is an important policy concern because the decision to work abroad depends not only on the attractiveness of foreign employment, but also on how workers trade location against income, career development, job security and working conditions. We revisit the stated job choices of Italian medical students studied by \citet{bertoni2026medical}. The experiment varies job location, income, job security, professional development, working conditions and skill match (\Cref{tab:medical_attributes}). It also provides a natural setting to examine how income sensitivity covaries with willingness to pay for switching other non-income job attributes, which implies a dependence structure that was difficult to estimate in the original analysis as informed by one of our coauthor in this article. 

We first benchmark four computational strategies using a WTP-space model that replicates the original model specification. It contains ten independent normally distributed WTP coefficients, including two alternative-specific constants, and a positive lognormally distributed income-sensitivity coefficient, giving 22 estimated parameters. Each strategy uses the same observed choices, 5,000 Halton draws, starting values and L-BFGS-B optimiser. Timings are from single optimisation runs on an Apple M1 Pro with 16 GB of memory, eight CPU threads and the MPS GPU backend. When we considered applying our strategies to an empirical application, access to AWS was unavailable due to a cap on the computational budget. We therefore performed the application on a personal laptop, again achieving reductions in runtime. This illustrates another practical feature of the prototype we developed, which is its support for various firmware. In addition to Intel, NVIDIA and AMD hardware, it supports Apple silicon chips, whose GPUs are readily accessible but are not often used by researchers studying health preferences. Because of the laptop's limited computational power, we were unable to benchmark FD based strategies for the full-correlation mixed logit model in this application. Nevertheless, we believe that the benchmark using the independent model provides sufficient evidence of the performance of AD-GPU-FP32.

\Cref{tab:medical_runtime} shows that FD-CPU-FP64 requires 811.49 seconds. Replacing FD with AD while retaining CPU execution and FP64 reduces runtime to 42.72 seconds, a 19 times speed-up. Under AD, moving to FP32 reduces runtime further to 25.75 seconds on the CPU and 24.58 seconds on the GPU, corresponding to speed-ups of 31.51 and 33.02, respectively. Like seen in previous experiments, AD accounts for most of the computational gain in this application. FP32 provides a further reduction in runtime, while GPU execution adds little beyond FP32 on this hardware. In addition, as explained above we only used AD-GPU-FP32 on full-correlation model which took us 29.89 seconds. The speed was only arbitrarily different with what has taken for independent model that potentially emphasize the power of AD. 

\begin{table}[htbp]
\centering\small
\caption{Attributes and levels in the medical job-choice experiment.}
\label{tab:medical_attributes}
\begin{tabular}{@{}>{\raggedright\arraybackslash}p{.34\textwidth}>{\raggedright\arraybackslash}p{\dimexpr.66\textwidth-2\tabcolsep\relax}@{}}
\toprule
Attribute & Levels \\
\midrule
Location & Italy \newline
Your favorite European foreign country \\
\addlinespace[.55em]
Annual net income adjusted by purchasing power parity & \texteuro{}20.000 \newline
\texteuro{}30.000 \newline
\texteuro{}40.000 \newline
\texteuro{}50.000 \newline
\texteuro{}60.000 \newline
\texteuro{}75.000 \\
\addlinespace[.55em]
Job security & 2-year fixed term contract, non-renewable \newline
2-year fixed term contract with 50\% renewal probability \newline
Permanent job \\
\addlinespace[.55em]
Professional development opportunities & Limited opportunities for research and training \newline
Some opportunities for research and training \newline
Good opportunities for research and training \\
\addlinespace[.55em]
Working conditions & High workload with frequent overtime work and night shifts \newline
Adequate workload with little overtime work and night shifts \\
\addlinespace[.55em]
Match of skills with job content & Your skills are higher than required by the job \newline
Your skills are exactly matched to what is required by the job \newline
Some of your skills are lower than required by the job and need further development \\
\bottomrule
\end{tabular}
\end{table}

\begin{table}[htbp]
\centering
\caption{Runtime benchmark for the replicated model.}
\label{tab:medical_runtime}
\small
\begin{tabular}{lrr}
\toprule
Strategy & Time (s)  & Speed-up \\
\midrule
FD-CPU-FP64 & 811.49  & 1.00 \\
AD-CPU-FP64 & 42.72  & 19.00 \\
AD-CPU-FP32 & 25.75  & 31.51 \\
AD-GPU-FP32 & 24.58  & 33.02 \\
\bottomrule
\end{tabular}
\par\vspace{.5em}
\begin{minipage}{0.96\textwidth}\footnotesize
\textit{Notes:} Speed-up is the FD-CPU-FP64 runtime divided the strategy's runtime.
\end{minipage}
\end{table}

\FloatBarrier

The replicated model closely matches the published results such that all coefficient signs and the substantive ordering of attribute means are preserved, with non-income attribute means differing by at most 8\% (\Cref{tab:medical_wtp_mean,tab:medical_wtp_sd}). The original article reports the number of simulation draws but ommited type or seed, so exact equality could not be achieved. Further allowing preference coefficients to correlate substantially improves model fit. Relative to the independent specification, AIC falls by 780.92 and BIC by 546.22 despite the addition of 55 parameters (\Cref{tab:medical_wtp_mean}). Together with the pattern of estimated correlations in \Cref{tab:medical_wtp_correlation}, the results show that the independence restriction was indeed empirically restrictive in modeling fitting, for this application.

Regardless of different correlation structures, the overall ordering of mean attribute valuations remains, while the magnitudes of both mean WTP and preference heterogeneity change appreciably. For example, mean WTP for some professional development opportunities increases from EUR~6,621 in the independent model to EUR~10,001 in the correlated model, while its standard deviation increases from EUR~7,613 to EUR~11,268. The richer covariance structure therefore changes both the mean and dispersion of several policy-relevant valuations beyond improving model fit.

The most important pattern concerns the relationship between location and income. Foreign-country WTP is positively correlated with income sensitivity. Respondents who are relatively more willing to work abroad are therefore also more responsive to income. This changes the interpretation of the negative average foreign-country WTP. The average respondent does not intrinsically prefer a foreign location, but the respondents who are more open to work abroad place greater weight on income. The combination points to a strong financial margin in the brain-drain decision: higher earnings can offset the compensation required to leave Italy, particularly among respondents whose choices are more income sensitive. In the meantime, foreign-country WTP was found to be strongly correlated negatively with WTP for some professional development opportunities. Respondents who are relatively more willing to work abroad therefore place less value on a moderate improvement in professional development. By contrast, the correlation with WTP for good development opportunities is much smaller. This distinction suggests that modest improvements in professional development are more closely aligned with the preferences of respondents inclined to remain in Italy, whereas strong development opportunities are valued more broadly across the sample. 

A similar pattern appears for short-term job security. Foreign-country WTP is negatively correlated with WTP for a fixed-term contract with a 50\% renewal probability. Respondents who place greater value on an increased probability of contract renewal are therefore less willing to work abroad. Incremental improvements in short-term employment security, like moderate improvements in professional development, are valued more strongly among respondents with a stronger preference for remaining in Italy. Permanent employment, however, shows a different relationship with working abroad. The correlation between foreign-country WTP and WTP for a permanent contract is small and statistically insignificant. This contrast indicates that incremental short-term security and permanent employment capture different preference dimensions. A permanent position represents a much longer-term commitment and may involve considerations such as migration arrangements, family commitments and country-specific attachment that are less relevant to a temporary increase in contract-renewal probability. The present experiment does not separately identify these mechanisms, but the absence of a relationship for permanent employment shows that the association between job security and working abroad is not uniform across forms of employment security.

Taken together, these results reveal a distinct preference profile associated with working abroad. Respondents who are relatively more willing to work abroad are more income sensitive, while placing less value on moderate professional-development opportunities and incremental short-term contract security. This pattern helps reconcile two features of the data that appear less informative when considered separately: medical students value remaining in Italy on average, yet working abroad can still be attractive when the financial return is sufficiently large. Improvements in domestic working conditions, career development and job security remain valuable on average, but their value is perceived differently across respondents, and some of these improvements are valued most strongly by those who were already more inclined to remain in Italy.

The full-correlation model could change what can be learned from the application. The independent specification identifies which job attributes are valued on average and how dispersed those valuations are; the correlated specification additionally identifies how these valuations combine within individuals. In the context of medical brain drain, this joint preference structure distinguishes a general preference for remaining in Italy from the financial and non-financial trade-offs associated with greater willingness to work abroad. The computational gains documented above thus do more than reduce estimation time but also makes it feasible to estimate preference structures that yield non-trivally different empirical evidence.

\begin{table}[p]
\centering\small
\caption{Mean parameters and model fit in the medical job-choice application.}
\label{tab:medical_wtp_mean}
\setlength{\tabcolsep}{4pt}
\renewcommand{\arraystretch}{1.04}
\begin{tabular}{@{}>{\raggedright\arraybackslash}p{.46\textwidth}*{3}{>{\raggedleft\arraybackslash}p{\dimexpr.18\textwidth-2\tabcolsep\relax}}@{}}
\toprule
Attribute / level & \shortstack[r]{Published\\independent} & \shortstack[r]{Replicated\\independent} & \shortstack[r]{Full\\correlation} \\
\midrule
\multicolumn{4}{@{}l}{\textbf{Location}} \\
\hspace*{.75em}Italy (reference) & --- & --- & --- \\
\hspace*{.75em}Favourite foreign European country & $-13{,}515^{***}$ & $-12{,}435^{***}$ & $-11{,}149^{***}$ \\
\addlinespace[.65em]
\multicolumn{4}{@{}l}{\textbf{Professional development opportunities}} \\
\hspace*{.75em}Limited opportunities (reference) & --- & --- & --- \\
\hspace*{.75em}Some opportunities & $6{,}572^{***}$ & $6{,}621^{***}$ & $10{,}001^{***}$ \\
\hspace*{.75em}Good opportunities & $10{,}998^{***}$ & $11{,}166^{***}$ & $13{,}146^{***}$ \\
\addlinespace[.65em]
\multicolumn{4}{@{}l}{\textbf{Job security}} \\
\hspace*{.75em}Fixed-term, non-renewable (reference) & --- & --- & --- \\
\hspace*{.75em}Fixed-term, 50\% renewal probability & $11{,}800^{***}$ & $12{,}261^{***}$ & $15{,}030^{***}$ \\
\hspace*{.75em}Permanent job & $8{,}878^{***}$ & $8{,}643^{***}$ & $11{,}807^{***}$ \\
\addlinespace[.65em]
\multicolumn{4}{@{}l}{\textbf{Working conditions}} \\
\hspace*{.75em}High workload (reference) & --- & --- & --- \\
\hspace*{.75em}Adequate workload & $12{,}740^{***}$ & $12{,}523^{***}$ & $14{,}220^{***}$ \\
\addlinespace[.65em]
\multicolumn{4}{@{}l}{\textbf{Match of skills with job content}} \\
\hspace*{.75em}Skills higher than required (reference) & --- & --- & --- \\
\hspace*{.75em}Skills exactly matched & $3{,}711^{***}$ & $3{,}696^{***}$ & $5{,}483^{***}$ \\
\hspace*{.75em}Skills lower than required & $3{,}002^{***}$ & $3{,}217^{***}$ & $4{,}197^{***}$ \\
\addlinespace[.65em]
\multicolumn{4}{@{}l}{\textbf{Annual net income (PPP adjusted)}} \\
\hspace*{.75em}Income sensitivity, $\alpha_i$ & $0.126^{***}$ & $0.119^{***}$ & $0.168^{***}$ \\
\addlinespace[.65em]
\multicolumn{4}{@{}l}{\textbf{Alternative-specific constants}} \\
\hspace*{.75em}Decline both offers (reference) & --- & --- & --- \\
\hspace*{.75em}Job A (ASC1) & $-37{,}686^{***}$ & $-38{,}134^{***}$ & $-38{,}691^{***}$ \\
\hspace*{.75em}Job B (ASC2) & $-39{,}961^{***}$ & $-40{,}200^{***}$ & $-42{,}253^{***}$ \\
\midrule
\multicolumn{4}{@{}l}{\textbf{Model fit}} \\
Parameters, $k$ & $22$ & $22$ & $77$ \\
Weighted simulated LL & \textit{n.r.} & $-5{,}763.23$ & $-5{,}317.77$ \\
AIC & \textit{n.r.} & $11{,}570.46$ & $10{,}789.54$ \\
BIC & \textit{n.r.} & $11{,}664.34$ & $11{,}118.12$ \\
\bottomrule
\end{tabular}
\par\vspace{.5em}
\begin{minipage}{\textwidth}\footnotesize
\textit{Notes:} The first column reproduces Table~2 of \citet{bertoni2026medical}, including its significance stars. WTP coefficients and constants are in euros per year; income sensitivity is per EUR~1,000. Income enters continuously with linear interpretation. Reference coefficients are fixed at zero (---). For model fit, $\mathrm{AIC}=-2\mathrm{LL}+2k$ and $\mathrm{BIC}=-2\mathrm{LL}+k\log(527)$. \textit{n.r.}: not reported in the published article. $^{*}p<0.10$, $^{**}p<0.05$, $^{***}p<0.01$.
\end{minipage}
\end{table}

\begin{table}[p]
\centering\small
\caption{Standard deviations of random coefficients in the medical job-choice application.}
\label{tab:medical_wtp_sd}
\setlength{\tabcolsep}{4pt}
\renewcommand{\arraystretch}{1.04}
\begin{tabular}{@{}>{\raggedright\arraybackslash}p{.46\textwidth}*{3}{>{\raggedleft\arraybackslash}p{\dimexpr.18\textwidth-2\tabcolsep\relax}}@{}}
\toprule
Attribute / level & \shortstack[r]{Published\\independent} & \shortstack[r]{Replicated\\independent} & \shortstack[r]{Full\\correlation} \\
\midrule
\multicolumn{4}{@{}l}{\textbf{Location}} \\
\hspace*{.75em}Italy (reference) & --- & --- & --- \\
\hspace*{.75em}Favourite foreign European country & $17{,}216^{***}$ & $17{,}643^{***}$ & $19{,}723^{***}$ \\
\addlinespace[.65em]
\multicolumn{4}{@{}l}{\textbf{Professional development opportunities}} \\
\hspace*{.75em}Limited opportunities (reference) & --- & --- & --- \\
\hspace*{.75em}Some opportunities & $8{,}182^{***}$ & $7{,}613^{***}$ & $11{,}268^{***}$ \\
\hspace*{.75em}Good opportunities & $7{,}641^{***}$ & $7{,}939^{***}$ & $12{,}451^{***}$ \\
\addlinespace[.65em]
\multicolumn{4}{@{}l}{\textbf{Job security}} \\
\hspace*{.75em}Fixed-term, non-renewable (reference) & --- & --- & --- \\
\hspace*{.75em}Fixed-term, 50\% renewal probability & $4{,}920^{***}$ & $5{,}205^{**}$ & $8{,}334^{***}$ \\
\hspace*{.75em}Permanent job & $13{,}071^{***}$ & $12{,}973^{***}$ & $15{,}219^{***}$ \\
\addlinespace[.65em]
\multicolumn{4}{@{}l}{\textbf{Working conditions}} \\
\hspace*{.75em}High workload (reference) & --- & --- & --- \\
\hspace*{.75em}Adequate workload & $8{,}408^{***}$ & $8{,}598^{***}$ & $9{,}981^{***}$ \\
\addlinespace[.65em]
\multicolumn{4}{@{}l}{\textbf{Match of skills with job content}} \\
\hspace*{.75em}Skills higher than required (reference) & --- & --- & --- \\
\hspace*{.75em}Skills exactly matched & $5{,}809^{***}$ & $6{,}828^{***}$ & $10{,}362^{***}$ \\
\hspace*{.75em}Skills lower than required & $8{,}849^{***}$ & $8{,}162^{***}$ & $9{,}223^{***}$ \\
\addlinespace[.65em]
\multicolumn{4}{@{}l}{\textbf{Annual net income (PPP adjusted)}} \\
\hspace*{.75em}Income sensitivity, $\alpha_i$ & $0.106^{***}$ & $0.092^{***}$ & $0.163^{***}$ \\
\addlinespace[.65em]
\multicolumn{4}{@{}l}{\textbf{Alternative-specific constants}} \\
\hspace*{.75em}Decline both offers (reference) & --- & --- & --- \\
\hspace*{.75em}Job A (ASC1) & $5{,}115^{***}$ & $4{,}676^{***}$ & $26{,}514^{***}$ \\
\hspace*{.75em}Job B (ASC2) & $2{,}931^{***}$ & $2{,}745^{***}$ & $25{,}556^{***}$ \\
\bottomrule
\end{tabular}
\par\vspace{.5em}
\begin{minipage}{\textwidth}\footnotesize
\textit{Notes:} The first column reproduces Table~2 of \citet{bertoni2026medical}, including its significance stars: $^{*}p<0.10$, $^{**}p<0.05$, $^{***}p<0.01$. 
\end{minipage}
\end{table}

\begin{table}[p]
\centering\footnotesize
\caption{Preference correlations in the full-correlation WTP model.}
\label{tab:medical_wtp_correlation}
\setlength{\tabcolsep}{2pt}
\renewcommand{\arraystretch}{1.3}
\begin{tabular*}{\textwidth}{@{\extracolsep{\fill}}l*{11}{r}@{}}
\toprule
 & (1) & (2) & (3) & (4) & (5) & (6) & (7) & (8) & (9) & (10) & (11) \\
\midrule
(1) & $1.00$ &  &  &  &  &  &  &  &  &  &  \\
(2) & $-0.53^{***}$ & $1.00$ &  &  &  &  &  &  &  &  &  \\
(3) & $-0.09^{*}$ & $0.74^{***}$ & $1.00$ &  &  &  &  &  &  &  &  \\
(4) & $-0.30^{**}$ & $0.22^{*}$ & $0.14$ & $1.00$ &  &  &  &  &  &  &  \\
(5) & $-0.06$ & $-0.26^{***}$ & $-0.22^{***}$ & $0.77^{***}$ & $1.00$ &  &  &  &  &  &  \\
(6) & $0.07$ & $0.18^{**}$ & $0.30^{***}$ & $-0.05$ & $-0.05$ & $1.00$ &  &  &  &  &  \\
(7) & $-0.08$ & $0.43^{***}$ & $0.58^{***}$ & $0.37^{**}$ & $0.11$ & $0.52^{***}$ & $1.00$ &  &  &  &  \\
(8) & $0.01$ & $0.05$ & $0.30^{***}$ & $0.51^{***}$ & $0.25^{***}$ & $0.11$ & $0.56^{***}$ & $1.00$ &  &  &  \\
(9) & $0.18^{***}$ & $-0.15^{*}$ & $-0.40^{***}$ & $-0.53^{***}$ & $-0.48^{***}$ & $-0.30^{***}$ & $-0.49^{***}$ & $-0.32^{***}$ & $1.00$ &  &  \\
(10) & $-0.39^{***}$ & $0.08$ & $-0.21^{***}$ & $-0.12$ & $-0.11^{*}$ & $-0.50^{***}$ & $-0.37^{***}$ & $-0.37^{***}$ & $0.29^{***}$ & $1.00$ &  \\
(11) & $-0.35^{***}$ & $0.05$ & $-0.21^{***}$ & $-0.18$ & $-0.13^{**}$ & $-0.47^{***}$ & $-0.45^{***}$ & $-0.43^{***}$ & $0.30^{***}$ & $0.99^{***}$ & $1.00$ \\
\bottomrule
\end{tabular*}
\par\vspace{.75em}
\begin{minipage}{\textwidth}\footnotesize
\textit{Coefficient key:} (1)~Foreign country; (2)~Some development; (3)~Good development; (4)~Fixed-term, 50\% renewal; (5)~Permanent job; (6)~Adequate workload; (7)~Skills exactly matched; (8)~Skills lower than required; (9)~Income sensitivity, $\alpha_i$; (10)~Job A constant; (11)~Job B constant.
\par\vspace{.5em}
\textit{Notes:} Lower-triangular population correlations of the ten normal WTP coefficients (including the two constants) and the positive lognormal income sensitivity $\alpha_i$. Income correlations are transformed from the fitted joint-normal distribution of WTP and $\log\alpha_i$. Reference coefficients are fixed at zero and have no defined correlations. $^{*}p<0.10$, $^{**}p<0.05$, $^{***}p<0.01$.
\end{minipage}
\end{table}

\FloatBarrier

%% file: sections/05_discussion.tex
\section{Discussion}
\label{sec:discussion}

Our controlled benchmark shows that AD substantially accelerates estimation of mixed logit, with additional gains from GPU execution and FP32 arithmetic. Overall, runtime was reduced significantly by using an AD-FP32-based strategy over conventional FD-FP64, while the reduction is even higher when using a GPU. As shown in the scaling experiments, these improvements expand as the number of individuals considered increases. More importantly, these runtime reductions were accompanied by broadly comparable parameter recovery among the strategies evaluated. The statistical robustness of the results was shown with 50 Monte Carlo repetitions by varying the data and simulation draws.

Each Monte Carlo trial in our results shows a non-trivial variance in parameter recovery, which implies different estimated models, which could further imply higher variance in secondary statistics such as choice probabilities and market uptake, which are usually used to inform policy. However, the mean statistics show that all strategies result in only trivial differences, and the secondary statistics should likewise show only trivial differences. This underpins the importance of multi-start and optimizing the impact of random noise, all of which require repeated runs and, in turn, underpin the fast estimation and significance of our study.

As the number of parameters increases, the cost of FD gradient construction rises because additional parameters require additional likelihood perturbations. AD does not exhibit the same parameter-by-parameter scaling in gradient computation, so its relative advantage becomes larger as model complexity increases. The empirical application provides complementary evidence from observed stated-preference data with more choice occasions for individuals. AD again accounts for most of the runtime reduction, while the incremental gain from GPU execution is smaller in that computing environment. The consistency of these qualitative patterns across the simulation benchmark and empirical application suggests that computational strategy can affect which mixed logit specifications can practically be estimated, which can, in turn, improve model fit to the data and provide richer empirical evidence. 

These gains arose naturally from our choices about how gradients are computed with FD and AD, the choice of firmware for computation, and the numerical precision used. This sequence of choices made has been shown to be crucial to the estimation time of mixed logit, whereby each part of the sequence accounts for different improvements. This study contributes to the literature on computational implementation in choice modelling by separating three computational dimensions that are often bundled together in software implementations \citep{arteaga2022xlogit, bierlaire2023biogeme, du2023torchchoice}. Our results have also shown that these computational choices could be interdependent such that the use of FP32 reduces the runtime more with AD-based strategies than FD-based strategies; the use of a GPU is also more beneficial when FP32 is used for precision. Nevertheless, from our study, we conclude that AD-FP32 strategies are computationally superior to their counterparts. Any researcher can benefit just by switching software alone without incurring additional monetary costs, such as those of purchasing a new device. 

The practical implications extend beyond reducing the runtime of a single estimation. Long estimation times can constrain the use of richer random-coefficient specifications, larger numbers of simulation draws, full covariance structures, multi-start optimisation, bootstrap inference, and extensive robustness analysis. Faster AD-FP32-based estimation relaxes these constraints and expands the range of empirical workflows that are computationally feasible. Larger draw counts become more practical when simulation noise is a concern \citep{bhat2003halton, hess2006mlhs, czajkowski2019simulation}, while the lower cost of repeated estimation can also make bootstrap inference \citep{hole2007bootstrap}, multi-start optimisation, and convergence checks more manageable.

Motivated by these findings and implications, we translate our research into more accessible software that could be used by empirical researchers. We provide an open-source prototype at
\url{https://huggingface.co/spaces/Wil2200/prefero} or \url{https://github.com/WilZ2200/prefero}. One could download it from GitHub, and we sincerely welcome all kinds of comments and are open to pull requests that make the software a better one to assist researchers. It is also hosted on Hugging Face, which provides free CPU and GPU cloud computing and stable servers. The software uses AD-FP32 as a computational strategy underpinned by PyTorch and would automatically use available firmware if GPUs from AMD, NVIDIA, or Apple are available. It is intended to reduce the coding required to implement AD- and GPU-compatible estimation for mixed logit and related discrete choice models such as generalised multinomial logit \citep{fiebig2010gmnl} and the latent class logit \citep{greene2003latent}. Our software supports flexible specification of the utility function by allowing interaction terms. It also allows a change of starting point and ordinary bootstrap estimation for inference.

Several limitations could as well suggest directions for future research. First, the relative performance of CPU and GPU execution is firmware- and implementation-dependent. GPU architecture, memory bandwidth, vectorisation, data-transfer costs, software backends, and the extent to which likelihood evaluation and optimisation remain on the GPU can all affect realised speed gains. Comparisons across a broader range of GPU architectures, CPU configurations, and software environments would potentially help establish the generalisability of the patterns being found in this research. However, we believe these are less significant compared with implementing better strategies. Second, the benchmark focuses on differentiable simulated maximum-likelihood estimation for mixed logit models with multivariate normal random coefficients. The relative advantages of AD, GPU execution, and FP32 arithmetic may differ for alternative mixing distributions, latent class models estimated with a two-stage optimizer such as expectation-maximization, Bayesian estimation, non-smooth likelihood components, or other estimation approaches. Extending the benchmark to these settings would help establish how broadly the observed computational patterns apply across choice-modelling methods.

%% file: sections/06_conclusion.tex
\section{Conclusion}
\label{sec:conclusion}

Using controlled Monte Carlo experiments, our study benchmarks gradient computation, processor choice and floating-point precision for mixed logit estimation. AD substantially reduces runtime even on CPUs, while combining AD with FP32 arithmetic and GPU execution delivers the largest gains, particularly as sample size and model complexity increase. Parameter recovery measured with MAE, MAPE and interval coverage, remains broadly comparable across strategies within each setting.

These computational savings make richer random-coefficient specifications, larger simulation draw counts, multi-start optimisation, bootstrap inference and robustness checks more feasible. By reducing the cost of both individual and repeated estimations, faster computational strategies expand researchers’ capacity to examine alternative specifications and assess the robustness of their findings. Computational choices therefore form part of the methodological practice supporting applied choice analysis.

Last, we provide a prototype software for discrete choice analysis at
\url{https://huggingface.co/spaces/Wil2200/prefero} or \url{https://github.com/WilZ2200/prefero} that use AD-FP32 strategies and adapt optimally to the computer used for analysis. 

%% file: sections/appendix_dgp_normalization.tex
\section{Proportionality Proof}
\label{app:dgp_normalization}

This appendix derives the proportionality used in \Cref{sec:simulated_panel}. For two alternatives \(j\) and \(k\) in the same choice task, define the attribute difference for person \(n\) in choice task \(t\) as
\[
    \Delta \bm{x}_{ntjk} = \bm{x}_{ntj}-\bm{x}_{ntk}.
\]
Let \(\bm{\beta}_n\) be the preference vector for that person. The deterministic utility difference is
\[
    V_{ntj}-V_{ntk}
    =
    \Delta \bm{x}_{ntjk}'\bm{\beta}_n.
\]

Because attributes are drawn independently across alternatives and dimensions from the same truncated normal distribution, if \(\sigma_x^2=\operatorname{Var}(x_{ntjm})\), then
\[
    \E[\Delta \bm{x}_{ntjk}]=\bm{0},
    \qquad
    \operatorname{Var}(\Delta \bm{x}_{ntjk})=2\sigma_x^2\bm{I}.
\]
Conditional on \(\bm{\beta}_n\),
\begin{align}
    \operatorname{Var}(V_{ntj}-V_{ntk}\mid \bm{\beta}_n)
    &=
    \operatorname{Var}(\Delta \bm{x}_{ntjk}'\bm{\beta}_n\mid \bm{\beta}_n) \nonumber\\
    &=
    \bm{\beta}_n'
    \operatorname{Var}(\Delta \bm{x}_{ntjk})
    \bm{\beta}_n \nonumber\\
    &=
    2\sigma_x^2\|\bm{\beta}_n\|^2.
    \label{eq:conditional_utility_variance}
\end{align}
The conditional mean is also zero:
\[
    \E[V_{ntj}-V_{ntk}\mid \bm{\beta}_n]
    =
    \E[\Delta \bm{x}_{ntjk}]'\bm{\beta}_n
    =
    0.
\]
Using the law of total variance,
\begin{align}
    \operatorname{Var}(V_{ntj}-V_{ntk})
    &=
    \E\!\left[
    \operatorname{Var}(V_{ntj}-V_{ntk}\mid \bm{\beta}_n)
    \right]
    +
    \operatorname{Var}\!\left(
    \E[V_{ntj}-V_{ntk}\mid \bm{\beta}_n]
    \right) \nonumber\\
    &=
    \E\!\left[
    \operatorname{Var}(V_{ntj}-V_{ntk}\mid \bm{\beta}_n)
    \right].
    \label{eq:total_variance_dgp}
\end{align}
Combining this with \Cref{eq:conditional_utility_variance} gives
\begin{equation}
    \operatorname{Var}(V_{ntj}-V_{ntk})
    =
    2\sigma_x^2\E[\|\bm{\beta}_n\|^2].
    \label{eq:utility_variance_second_moment}
\end{equation}

The remaining term is the second moment of \(\bm{\beta}_n\). Since
\[
    \bm{\beta}_n=\bm{\mu}_0+\bm{L}_0\bm{z}_n,
    \qquad
    \bm{z}_n\sim\mathcal{N}(\bm{0},\bm{I}),
\]
we have
\begin{align}
    \E[\|\bm{\beta}_n\|^2]
    &=
    \E[(\bm{\mu}_0+\bm{L}_0\bm{z}_n)'(\bm{\mu}_0+\bm{L}_0\bm{z}_n)] \nonumber\\
    &=
    \|\bm{\mu}_0\|^2+
    \operatorname{tr}(\bm{L}_0\bm{L}_0').
    \label{eq:beta_second_moment_appendix}
\end{align}
Combining \Cref{eq:utility_variance_second_moment,eq:beta_second_moment_appendix},
\begin{equation}
    \operatorname{Var}(V_{ntj}-V_{ntk})
    =
    2\sigma_x^2
    \left(
    \|\bm{\mu}_0\|^2+
    \operatorname{tr}(\bm{L}_0\bm{L}_0')
    \right).
    \label{eq:utility_diff_variance}
\end{equation}
Since \(2\sigma_x^2\) is fixed by the attribute distribution and does not vary with \(p_{\mathrm{mean}}\),
\begin{equation}
    \operatorname{Var}(V_{ntj}-V_{ntk})
    \propto
    \|\bm{\mu}_0\|^2+
    \operatorname{tr}(\bm{L}_0\bm{L}_0').
    \label{eq:utility_variance_proportional}
\end{equation}

The constant \(\sigma_x^2\) is the marginal variance of the simulated attribute distribution \(x_{ntjm}\sim\mathcal{TN}_{[0,3]}(1.5,0.9^2)\). For a normal distribution with mean \(\mu_x\) and standard deviation \(s_x\), truncated to \([a,b]\), let
\[
    \alpha=\frac{a-\mu_x}{s_x},
    \qquad
    \gamma=\frac{b-\mu_x}{s_x}.
\]
If \(\phi(\cdot)\) and \(\Phi(\cdot)\) denote the standard normal probability density function and cumulative distribution function, respectively, the truncated variance is
\begin{align}
    \sigma_x^2
    =
    s_x^2
    \left[
    1+
    \frac{\alpha\phi(\alpha)-\gamma\phi(\gamma)}
    {\Phi(\gamma)-\Phi(\alpha)}
    -
    \left(
    \frac{\phi(\alpha)-\phi(\gamma)}
    {\Phi(\gamma)-\Phi(\alpha)}
    \right)^2
    \right].
    \label{eq:truncated_attribute_variance}
\end{align}
In our DGP, \(\mu_x=1.5\), \(s_x=0.9\), \(a=0\), and \(b=3\), so \(\alpha=-5/3\), \(\gamma=5/3\), and \(\sigma_x^2\approx 0.513\). Thus, under these simulated attribute draws, fixing \(\|\bm{\mu}_0\|^2+\operatorname{tr}(\bm{L}_0\bm{L}_0')\) keeps the deterministic utility scale comparable as dimension changes.

%% file: sections/appendix_runtime_tables.tex
\section{Detailed Runtime Scaling Results}
\label{app:runtime_tables}

\renewcommand{\thetable}{C\arabic{table}}
\renewcommand{\theHtable}{C\arabic{table}}
\setcounter{table}{0}

This appendix reports the full runtime values for the two scaling experiments. In \Cref{tab:app_n_scaling_runtime}, the number of respondents \(N\) varies while the number of estimated parameters is fixed at \(p=20\). In \Cref{tab:app_p_scaling_runtime}, \(p\) varies while \(N=1{,}000\). All entries summarise 50 Monte Carlo repetitions. Runtime is wall-clock estimation time in seconds. For a strategy \(m\) in design cell \(c\), speed-up is computed within each repetition as
\[
    \mathrm{Speed\mbox{-}up}_{m,c}
    =
    \frac{T_{\mathrm{FD\mbox{-}CPU\mbox{-}FP64},c}}
         {T_{m,c}},
\]
where \(T_{\mathrm{FD\mbox{-}CPU\mbox{-}FP64},c}\) is the runtime of the reference strategy in the same design cell. We report the geometric mean of speed-up across repetitions; values greater than one indicate faster estimation than the reference strategy. Parentheses report sample standard deviations across repetitions. Dashes in \Cref{tab:app_p_scaling_runtime} represent experiment cells omitted due to high computational cost.

\begingroup
\small
\setlength{\tabcolsep}{5pt}
\begin{longtable}{@{}>{\raggedleft\arraybackslash}p{.09\linewidth}>{\raggedright\arraybackslash}p{.35\linewidth}>{\raggedleft\arraybackslash}p{.26\linewidth}>{\raggedleft\arraybackslash}p{\dimexpr.30\linewidth-6\tabcolsep\relax}@{}}
\caption{\(N\)-scaling runtime and speed-up (\(p=20\); 50 Monte Carlo repetitions).}
\label{tab:app_n_scaling_runtime}\\
\toprule
\(N\) & Strategy & Runtime (s) & Speed-up \\
\midrule
\endfirsthead
\toprule
\(N\) & Strategy & Runtime (s) & Speed-up \\
\midrule
\endhead
\bottomrule
\endfoot
200 & FD-CPU-FP64 (ref.) & 58.4 (11.5) & 1.00 (0.00) \\
200 & AD-CPU-FP64 & 6.6 (1.2) & 8.85 (0.46) \\
200 & AD-CPU-FP32 & 1.8 (0.3) & 32.32 (6.79) \\
200 & AD-GPU-FP32 & 0.7 (0.1) & 88.62 (18.15) \\
\addlinespace
500 & FD-CPU-FP64 (ref.) & 167.0 (39.8) & 1.00 (0.00) \\
500 & AD-CPU-FP64 & 17.4 (3.7) & 9.56 (0.83) \\
500 & AD-CPU-FP32 & 4.1 (0.9) & 40.51 (10.79) \\
500 & AD-GPU-FP32 & 1.0 (0.2) & 170.53 (42.25) \\
\addlinespace
1,000 & FD-CPU-FP64 (ref.) & 331.2 (101.8) & 1.00 (0.00) \\
1,000 & AD-CPU-FP64 & 44.9 (13.5) & 7.35 (0.50) \\
1,000 & AD-CPU-FP32 & 7.7 (2.0) & 42.56 (12.69) \\
1,000 & AD-GPU-FP32 & 1.4 (0.3) & 226.19 (65.36) \\
\addlinespace
2,000 & FD-CPU-FP64 (ref.) & 705.3 (268.4) & 1.00 (0.00) \\
2,000 & AD-CPU-FP64 & 79.3 (27.5) & 8.82 (0.55) \\
2,000 & AD-CPU-FP32 & 17.0 (4.1) & 40.35 (9.89) \\
2,000 & AD-GPU-FP32 & 2.3 (0.6) & 301.47 (64.27) \\
\addlinespace
5,000 & FD-CPU-FP64 (ref.) & 1,737.8 (490.3) & 1.00 (0.00) \\
5,000 & AD-CPU-FP64 & 170.4 (44.2) & 10.17 (0.36) \\
5,000 & AD-CPU-FP32 & 46.9 (11.3) & 36.87 (8.35) \\
5,000 & AD-GPU-FP32 & 4.1 (0.8) & 416.31 (96.99) \\
\addlinespace
10,000 & FD-CPU-FP64 (ref.) & 3,370.9 (215.8) & 1.00 (0.00) \\
10,000 & AD-CPU-FP64 & 270.9 (17.3) & 12.44 (0.04) \\
10,000 & AD-CPU-FP32 & 79.5 (13.9) & 42.99 (8.53) \\
10,000 & AD-GPU-FP32 & 6.9 (1.0) & 493.75 (77.02) \\
\end{longtable}
\endgroup

\footnotesize\textit{Note:} Runtimes are means across the 50 Monte Carlo repetitions, with sample standard deviations in parentheses. Speed-ups are geometric means of within-repetition ratios relative to FD-CPU-FP64 within the same \(N\) cell, with sample standard deviations of those ratios in parentheses.
\normalsize

\begingroup
\small
\setlength{\tabcolsep}{5pt}
\begin{longtable}{@{}>{\raggedleft\arraybackslash}p{.09\linewidth}>{\raggedright\arraybackslash}p{.35\linewidth}>{\raggedleft\arraybackslash}p{.26\linewidth}>{\raggedleft\arraybackslash}p{\dimexpr.30\linewidth-6\tabcolsep\relax}@{}}
\caption{\(p\)-scaling runtime and speed-up (\(N=1{,}000\); 50 Monte Carlo repetitions).}
\label{tab:app_p_scaling_runtime}\\
\toprule
\(p\) & Strategy & Runtime (s) & Speed-up \\
\midrule
\endfirsthead
\toprule
\(p\) & Strategy & Runtime (s) & Speed-up \\
\midrule
\endhead
\bottomrule
\endfoot
20 & FD-CPU-FP64 (ref.) & 331.2 (101.8) & 1.00 (0.00) \\
20 & AD-CPU-FP64 & 45.0 (13.5) & 7.34 (0.51) \\
20 & AD-CPU-FP32 & 7.7 (2.0) & 42.61 (12.75) \\
20 & AD-GPU-FP32 & 1.5 (0.4) & 219.28 (64.53) \\
\addlinespace
65 & FD-CPU-FP64 (ref.) & 3,718.9 (649.7) & 1.00 (0.00) \\
65 & AD-CPU-FP64 & 142.7 (24.6) & 26.06 (1.75) \\
65 & AD-CPU-FP32 & 20.3 (3.5) & 183.17 (52.55) \\
65 & AD-GPU-FP32 & 3.2 (0.5) & 1,147.4 (327.6) \\
\addlinespace
135 & AD-CPU-FP64 & 313.9 (64.4) & -- \\
135 & AD-CPU-FP32 & 33.4 (6.8) & -- \\
135 & AD-GPU-FP32 & 5.6 (1.1) & -- \\
\addlinespace
230 & AD-CPU-FP64 & 537.6 (75.0) & -- \\
230 & AD-CPU-FP32 & 73.6 (16.7) & -- \\
230 & AD-GPU-FP32 & 10.3 (2.4) & -- \\
\end{longtable}
\endgroup

\footnotesize\textit{Note:} Runtimes are means across the 50 Monte Carlo repetitions, with sample standard deviations in parentheses. Speed-ups are geometric means of within-repetition ratios relative to FD-CPU-FP64 within the same \(p\) cell, with sample standard deviations of those ratios in parentheses. Speed-ups are omitted for \(p=135\) and \(p=230\) because FD-CPU-FP64 was not estimated for these model sizes due to high computational cost.
\normalsize

%% file: sections/appendix_recovery_tables.tex
\begin{landscape}
\section{Detailed Parameter Recovery Results}
\label{app:recovery_tables}
\renewcommand{\thetable}{D\arabic{table}}
\renewcommand{\theHtable}{D\arabic{table}}
\setcounter{table}{0}

This appendix reports parameter recovery for the main benchmark and the two scaling experiments. The parameter blocks are the mean vector \(\bm{\mu}\) and the unique covariance elements \(\operatorname{vech}(\bm{\Omega})\), where \(\bm{\Omega}=\bm{L}\bm{L}'\). For a block \(b\) with \(K_b\) elements in repetition \(s\),
\[
\operatorname{MAE}_{s,b}=\frac{1}{K_b}\sum_{k=1}^{K_b}
\left|\widehat{\theta}_{s,b,k}-\theta_{0,b,k}\right|,
\qquad
\operatorname{MAPE}_{s,b}=\frac{100}{K_b}\sum_{k=1}^{K_b}
\frac{\left|\widehat{\theta}_{s,b,k}-\theta_{0,b,k}\right|}{\left|\theta_{0,b,k}\right|}.
\]
All true elements in these blocks are nonzero. MAPE is sensitive to small true values, including the off-diagonal covariances, and is reported alongside MAE. Coverage uses approximate 95 percent BHHH intervals \citep{berndt1974bhhh}, with the delta method for covariance elements. Metrics are computed within each repetition and then averaged across 50 repetitions; parentheses report sample standard deviations across repetitions.

\begingroup
\small
\setlength{\tabcolsep}{5pt}
\begin{longtable}{@{}>{\raggedright\arraybackslash}p{0.19\linewidth}>{\raggedright\arraybackslash}p{0.20\linewidth}>{\raggedleft\arraybackslash}p{0.21\linewidth}>{\raggedleft\arraybackslash}p{0.19\linewidth}>{\raggedleft\arraybackslash}p{0.14\linewidth}@{}}
\caption{Main benchmark parameter recovery ($N=1{,}000$, $P=20$; 50 Monte Carlo repetitions).}
\label{tab:app_baseline_recovery}\\
\toprule
Strategy & Parameter block & MAE & MAPE (\%) & Coverage rate (\%) \\
\midrule
\endfirsthead
\toprule
Strategy & Parameter block & MAE & MAPE (\%) & Coverage rate (\%) \\
\midrule
\endhead
\bottomrule
\endfoot
FD-CPU-FP64 (ref.) & $\bm{\mu}$ & 0.01868 (0.00701) & 4.82 (1.81) & 94.0 (12.9) \\*
FD-CPU-FP64 (ref.) & $\operatorname{vech}(\bm{\Omega})$ & 0.01466 (0.00279) & 64.34 (13.37) & 95.6 (5.3) \\
\addlinespace
AD-CPU-FP64 & $\bm{\mu}$ & 0.01868 (0.00701) & 4.82 (1.81) & 94.0 (12.9) \\*
AD-CPU-FP64 & $\operatorname{vech}(\bm{\Omega})$ & 0.01466 (0.00279) & 64.34 (13.37) & 95.6 (5.3) \\
\addlinespace
AD-CPU-FP32 & $\bm{\mu}$ & 0.01868 (0.00701) & 4.82 (1.81) & 94.0 (12.9) \\*
AD-CPU-FP32 & $\operatorname{vech}(\bm{\Omega})$ & 0.01469 (0.00278) & 64.52 (13.40) & 95.9 (5.4) \\
\addlinespace
AD-GPU-FP32 & $\bm{\mu}$ & 0.01868 (0.00701) & 4.82 (1.81) & 94.0 (12.9) \\*
AD-GPU-FP32 & $\operatorname{vech}(\bm{\Omega})$ & 0.01469 (0.00279) & 64.51 (13.44) & 95.7 (5.3) \\
\addlinespace
FD-CPU-FP32 & $\bm{\mu}$ & 0.01869 (0.00701) & 4.83 (1.81) & 94.0 (12.9) \\*
FD-CPU-FP32 & $\operatorname{vech}(\bm{\Omega})$ & 0.01465 (0.00280) & 64.24 (13.56) & 95.9 (5.4) \\
\addlinespace
FD-GPU-FP32 & $\bm{\mu}$ & 0.01866 (0.00700) & 4.82 (1.81) & 94.0 (12.9) \\*
FD-GPU-FP32 & $\operatorname{vech}(\bm{\Omega})$ & 0.01468 (0.00280) & 64.38 (13.45) & 96.1 (5.4) \\
\end{longtable}
\endgroup

\footnotesize\textit{Note:} Entries are means across 50 repetitions, with sample standard deviations in parentheses. MAE and MAPE are computed across parameter elements within each repetition. MAPE divides each absolute error by the absolute true parameter value and is expressed as a percentage. Coverage is the percentage of parameter elements whose true value lies inside the 95 percent BHHH interval; covariance intervals use the delta method.
\normalsize

\begingroup
\small
\setlength{\tabcolsep}{5pt}
\begin{longtable}{@{}>{\raggedleft\arraybackslash}p{0.05\linewidth}>{\raggedright\arraybackslash}p{0.18\linewidth}>{\raggedright\arraybackslash}p{0.17\linewidth}>{\raggedleft\arraybackslash}p{0.19\linewidth}>{\raggedleft\arraybackslash}p{0.17\linewidth}>{\raggedleft\arraybackslash}p{0.16\linewidth}@{}}
\caption{\(N\)-scaling parameter recovery (\(P=20\); 50 Monte Carlo repetitions).}
\label{tab:app_n_scaling_recovery}\\
\toprule
\(N\) & Strategy & Parameter block & MAE & MAPE (\%) & Coverage rate (\%) \\
\midrule
\endfirsthead
\toprule
\(N\) & Strategy & Parameter block & MAE & MAPE (\%) & Coverage rate (\%) \\
\midrule
\endhead
\bottomrule
\endfoot
200 & FD-CPU-FP64 (ref.) & $\bm{\mu}$ & 0.04134 (0.01443) & 10.67 (3.73) & 98.4 (5.5) \\*
200 & FD-CPU-FP64 (ref.) & $\operatorname{vech}(\bm{\Omega})$ & 0.02889 (0.00793) & 122.35 (32.18) & 97.9 (5.5) \\
200 & AD-CPU-FP64 & $\bm{\mu}$ & 0.04134 (0.01443) & 10.67 (3.73) & 98.4 (5.5) \\*
200 & AD-CPU-FP64 & $\operatorname{vech}(\bm{\Omega})$ & 0.02889 (0.00793) & 122.34 (32.19) & 97.9 (5.5) \\
200 & AD-CPU-FP32 & $\bm{\mu}$ & 0.04128 (0.01441) & 10.66 (3.72) & 98.4 (5.5) \\*
200 & AD-CPU-FP32 & $\operatorname{vech}(\bm{\Omega})$ & 0.02893 (0.00791) & 122.56 (32.02) & 98.3 (5.2) \\
200 & AD-GPU-FP32 & $\bm{\mu}$ & 0.04128 (0.01441) & 10.66 (3.72) & 98.4 (5.5) \\*
200 & AD-GPU-FP32 & $\operatorname{vech}(\bm{\Omega})$ & 0.02892 (0.00788) & 122.60 (31.83) & 98.3 (5.2) \\
\addlinespace
500 & FD-CPU-FP64 (ref.) & $\bm{\mu}$ & 0.02567 (0.00817) & 6.63 (2.11) & 98.0 (6.1) \\*
500 & FD-CPU-FP64 (ref.) & $\operatorname{vech}(\bm{\Omega})$ & 0.01823 (0.00367) & 77.38 (16.29) & 96.9 (5.1) \\
500 & AD-CPU-FP64 & $\bm{\mu}$ & 0.02567 (0.00817) & 6.63 (2.11) & 98.0 (6.1) \\*
500 & AD-CPU-FP64 & $\operatorname{vech}(\bm{\Omega})$ & 0.01823 (0.00367) & 77.38 (16.29) & 96.9 (5.1) \\
500 & AD-CPU-FP32 & $\bm{\mu}$ & 0.02568 (0.00815) & 6.63 (2.10) & 98.0 (6.1) \\*
500 & AD-CPU-FP32 & $\operatorname{vech}(\bm{\Omega})$ & 0.01821 (0.00371) & 77.27 (16.47) & 96.9 (4.9) \\
500 & AD-GPU-FP32 & $\bm{\mu}$ & 0.02567 (0.00815) & 6.63 (2.11) & 98.0 (6.1) \\*
500 & AD-GPU-FP32 & $\operatorname{vech}(\bm{\Omega})$ & 0.01825 (0.00370) & 77.46 (16.31) & 96.8 (4.7) \\
\addlinespace
1,000 & FD-CPU-FP64 (ref.) & $\bm{\mu}$ & 0.01868 (0.00701) & 4.82 (1.81) & 94.0 (12.9) \\*
1,000 & FD-CPU-FP64 (ref.) & $\operatorname{vech}(\bm{\Omega})$ & 0.01466 (0.00279) & 64.34 (13.37) & 95.6 (5.3) \\
1,000 & AD-CPU-FP64 & $\bm{\mu}$ & 0.01868 (0.00701) & 4.82 (1.81) & 94.0 (12.9) \\*
1,000 & AD-CPU-FP64 & $\operatorname{vech}(\bm{\Omega})$ & 0.01466 (0.00279) & 64.34 (13.37) & 95.6 (5.3) \\
1,000 & AD-CPU-FP32 & $\bm{\mu}$ & 0.01868 (0.00701) & 4.82 (1.81) & 94.0 (12.9) \\*
1,000 & AD-CPU-FP32 & $\operatorname{vech}(\bm{\Omega})$ & 0.01469 (0.00278) & 64.52 (13.40) & 95.9 (5.4) \\
1,000 & AD-GPU-FP32 & $\bm{\mu}$ & 0.01868 (0.00701) & 4.82 (1.81) & 94.0 (12.9) \\*
1,000 & AD-GPU-FP32 & $\operatorname{vech}(\bm{\Omega})$ & 0.01468 (0.00278) & 64.48 (13.44) & 95.7 (5.3) \\
\addlinespace
2,000 & FD-CPU-FP64 (ref.) & $\bm{\mu}$ & 0.01233 (0.00436) & 3.18 (1.13) & 95.2 (10.3) \\*
2,000 & FD-CPU-FP64 (ref.) & $\operatorname{vech}(\bm{\Omega})$ & 0.01036 (0.00181) & 44.69 (8.33) & 94.9 (5.3) \\
2,000 & AD-CPU-FP64 & $\bm{\mu}$ & 0.01233 (0.00436) & 3.18 (1.13) & 95.2 (10.3) \\*
2,000 & AD-CPU-FP64 & $\operatorname{vech}(\bm{\Omega})$ & 0.01036 (0.00181) & 44.69 (8.33) & 94.9 (5.3) \\
2,000 & AD-CPU-FP32 & $\bm{\mu}$ & 0.01234 (0.00437) & 3.19 (1.13) & 95.2 (10.3) \\*
2,000 & AD-CPU-FP32 & $\operatorname{vech}(\bm{\Omega})$ & 0.01037 (0.00181) & 44.70 (8.29) & 95.2 (5.2) \\
2,000 & AD-GPU-FP32 & $\bm{\mu}$ & 0.01234 (0.00437) & 3.19 (1.13) & 95.2 (10.3) \\*
2,000 & AD-GPU-FP32 & $\operatorname{vech}(\bm{\Omega})$ & 0.01037 (0.00181) & 44.71 (8.30) & 95.1 (5.2) \\
\addlinespace
5,000 & FD-CPU-FP64 (ref.) & $\bm{\mu}$ & 0.00808 (0.00280) & 2.08 (0.72) & 96.0 (8.1) \\*
5,000 & FD-CPU-FP64 (ref.) & $\operatorname{vech}(\bm{\Omega})$ & 0.00709 (0.00160) & 31.08 (7.26) & 93.1 (7.5) \\
5,000 & AD-CPU-FP64 & $\bm{\mu}$ & 0.00808 (0.00280) & 2.08 (0.72) & 96.0 (8.1) \\*
5,000 & AD-CPU-FP64 & $\operatorname{vech}(\bm{\Omega})$ & 0.00709 (0.00160) & 31.08 (7.26) & 93.1 (7.5) \\
5,000 & AD-CPU-FP32 & $\bm{\mu}$ & 0.00807 (0.00279) & 2.08 (0.72) & 96.0 (8.1) \\*
5,000 & AD-CPU-FP32 & $\operatorname{vech}(\bm{\Omega})$ & 0.00706 (0.00159) & 30.91 (7.21) & 93.1 (7.5) \\
5,000 & AD-GPU-FP32 & $\bm{\mu}$ & 0.00807 (0.00280) & 2.08 (0.72) & 96.4 (7.8) \\*
5,000 & AD-GPU-FP32 & $\operatorname{vech}(\bm{\Omega})$ & 0.00703 (0.00157) & 30.78 (7.08) & 93.2 (7.3) \\
\addlinespace
10,000 & FD-CPU-FP64 (ref.) & $\bm{\mu}$ & 0.00624 (0.00204) & 1.61 (0.53) & 93.2 (11.1) \\*
10,000 & FD-CPU-FP64 (ref.) & $\operatorname{vech}(\bm{\Omega})$ & 0.00492 (0.00103) & 20.96 (3.93) & 95.1 (6.0) \\
10,000 & AD-CPU-FP64 & $\bm{\mu}$ & 0.00624 (0.00204) & 1.61 (0.53) & 93.2 (11.1) \\*
10,000 & AD-CPU-FP64 & $\operatorname{vech}(\bm{\Omega})$ & 0.00492 (0.00103) & 20.96 (3.93) & 95.1 (6.0) \\
10,000 & AD-CPU-FP32 & $\bm{\mu}$ & 0.00623 (0.00204) & 1.61 (0.53) & 93.2 (11.1) \\*
10,000 & AD-CPU-FP32 & $\operatorname{vech}(\bm{\Omega})$ & 0.00491 (0.00105) & 20.92 (4.01) & 95.2 (5.6) \\
10,000 & AD-GPU-FP32 & $\bm{\mu}$ & 0.00625 (0.00204) & 1.61 (0.53) & 93.2 (11.1) \\*
10,000 & AD-GPU-FP32 & $\operatorname{vech}(\bm{\Omega})$ & 0.00491 (0.00104) & 20.91 (3.99) & 95.2 (5.4) \\
\end{longtable}
\endgroup

\footnotesize\textit{Note:} Entries are means across 50 repetitions, with sample standard deviations in parentheses. MAE and MAPE are computed across parameter elements within each repetition. MAPE divides each absolute error by the absolute true parameter value and is expressed as a percentage. Coverage is the percentage of parameter elements whose true value lies inside the 95 percent BHHH interval; covariance intervals use the delta method.
\normalsize

\clearpage
\begingroup
\small
\setlength{\tabcolsep}{5pt}
\begin{longtable}{@{}>{\raggedleft\arraybackslash}p{0.05\linewidth}>{\raggedright\arraybackslash}p{0.18\linewidth}>{\raggedright\arraybackslash}p{0.17\linewidth}>{\raggedleft\arraybackslash}p{0.19\linewidth}>{\raggedleft\arraybackslash}p{0.17\linewidth}>{\raggedleft\arraybackslash}p{0.16\linewidth}@{}}
\caption{\(P\)-scaling parameter recovery (\(N=1{,}000\); 50 Monte Carlo repetitions).}
\label{tab:app_p_scaling_recovery}\\
\toprule
\(P\) & Strategy & Parameter block & MAE & MAPE (\%) & Coverage rate (\%) \\
\midrule
\endfirsthead
\toprule
\(P\) & Strategy & Parameter block & MAE & MAPE (\%) & Coverage rate (\%) \\
\midrule
\endhead
\bottomrule
\endfoot
20 & FD-CPU-FP64 (ref.) & $\bm{\mu}$ & 0.01868 (0.00701) & 4.82 (1.81) & 94.0 (12.9) \\*
20 & FD-CPU-FP64 (ref.) & $\operatorname{vech}(\bm{\Omega})$ & 0.01466 (0.00279) & 64.34 (13.37) & 95.6 (5.3) \\
20 & AD-CPU-FP64 & $\bm{\mu}$ & 0.01868 (0.00701) & 4.82 (1.81) & 94.0 (12.9) \\*
20 & AD-CPU-FP64 & $\operatorname{vech}(\bm{\Omega})$ & 0.01466 (0.00279) & 64.34 (13.37) & 95.6 (5.3) \\
20 & AD-CPU-FP32 & $\bm{\mu}$ & 0.01868 (0.00701) & 4.82 (1.81) & 94.0 (12.9) \\*
20 & AD-CPU-FP32 & $\operatorname{vech}(\bm{\Omega})$ & 0.01469 (0.00278) & 64.52 (13.40) & 95.9 (5.4) \\
20 & AD-GPU-FP32 & $\bm{\mu}$ & 0.01867 (0.00701) & 4.82 (1.81) & 94.0 (12.9) \\*
20 & AD-GPU-FP32 & $\operatorname{vech}(\bm{\Omega})$ & 0.01469 (0.00279) & 64.51 (13.44) & 95.7 (5.3) \\
\addlinespace
65 & FD-CPU-FP64 (ref.) & $\bm{\mu}$ & 0.01878 (0.00547) & 6.86 (2.00) & 95.4 (7.6) \\*
65 & FD-CPU-FP64 (ref.) & $\operatorname{vech}(\bm{\Omega})$ & 0.01103 (0.00196) & 107.33 (16.77) & 98.4 (1.7) \\
65 & AD-CPU-FP64 & $\bm{\mu}$ & 0.01878 (0.00547) & 6.86 (2.00) & 95.4 (7.6) \\*
65 & AD-CPU-FP64 & $\operatorname{vech}(\bm{\Omega})$ & 0.01103 (0.00196) & 107.33 (16.77) & 98.4 (1.7) \\
65 & AD-CPU-FP32 & $\bm{\mu}$ & 0.01877 (0.00551) & 6.85 (2.01) & 95.6 (7.6) \\*
65 & AD-CPU-FP32 & $\operatorname{vech}(\bm{\Omega})$ & 0.01099 (0.00197) & 107.00 (16.82) & 98.4 (1.7) \\
65 & AD-GPU-FP32 & $\bm{\mu}$ & 0.01878 (0.00552) & 6.86 (2.01) & 95.6 (7.6) \\*
65 & AD-GPU-FP32 & $\operatorname{vech}(\bm{\Omega})$ & 0.01101 (0.00197) & 107.15 (16.87) & 98.4 (1.7) \\
\addlinespace
\pagebreak
135 & AD-CPU-FP64 & $\bm{\mu}$ & 0.02290 (0.00507) & 10.24 (2.27) & 93.7 (7.4) \\*
135 & AD-CPU-FP64 & $\operatorname{vech}(\bm{\Omega})$ & 0.01165 (0.00154) & 170.15 (20.04) & 98.8 (1.1) \\
135 & AD-CPU-FP32 & $\bm{\mu}$ & 0.02281 (0.00502) & 10.20 (2.24) & 93.7 (7.2) \\*
135 & AD-CPU-FP32 & $\operatorname{vech}(\bm{\Omega})$ & 0.01160 (0.00154) & 169.84 (20.15) & 98.8 (1.1) \\
135 & AD-GPU-FP32 & $\bm{\mu}$ & 0.02281 (0.00502) & 10.20 (2.25) & 93.7 (7.2) \\*
135 & AD-GPU-FP32 & $\operatorname{vech}(\bm{\Omega})$ & 0.01161 (0.00154) & 169.87 (20.14) & 98.8 (1.1) \\
\addlinespace
230 & AD-CPU-FP64 & $\bm{\mu}$ & 0.02514 (0.00489) & 12.98 (2.53) & 92.0 (7.4) \\*
230 & AD-CPU-FP64 & $\operatorname{vech}(\bm{\Omega})$ & 0.01308 (0.00154) & 254.96 (28.35) & 98.7 (0.8) \\
230 & AD-CPU-FP32 & $\bm{\mu}$ & 0.02495 (0.00485) & 12.88 (2.50) & 92.5 (7.1) \\*
230 & AD-CPU-FP32 & $\operatorname{vech}(\bm{\Omega})$ & 0.01298 (0.00153) & 253.36 (28.15) & 98.8 (0.8) \\
230 & AD-GPU-FP32 & $\bm{\mu}$ & 0.02497 (0.00485) & 12.89 (2.50) & 92.5 (7.1) \\*
230 & AD-GPU-FP32 & $\operatorname{vech}(\bm{\Omega})$ & 0.01299 (0.00153) & 253.47 (28.25) & 98.8 (0.8) \\
\end{longtable}
\endgroup

\footnotesize\textit{Note:} Entries are means across 50 repetitions, with sample standard deviations in parentheses. MAE and MAPE are computed across parameter elements within each repetition. MAPE divides each absolute error by the absolute true parameter value and is expressed as a percentage. Coverage is the percentage of parameter elements whose true value lies inside the 95 percent BHHH interval; covariance intervals use the delta method. FD-CPU-FP64 was omitted for \(P=135\) and \(P=230\) due to high computational cost.
\normalsize
\end{landscape}

%% file: references.bib
@article{nouwens2025evolving,
  title={The Evolving Landscape of Discrete Choice Experiments in Health Economics: A Systematic Review: SPH Nouwens et al.},
  author={Nouwens, Sven Petrus Henricus and Marceta, Stella Maria and Bui, Michael and van Dijk, Daisy Maria Alberta Hendrika and Groothuis-Oudshoorn, Catharina Gerarda Maria and Veldwijk, Jorien and van Til, Janine Astrid and de Bekker-Grob, Esther Wilhelmina},
  journal={Pharmacoeconomics},
  volume={43},
  number={8},
  pages={879--936},
  year={2025},
  publisher={Springer}
}

@article{liu2025exploring,
  title={Exploring the potential of computational graph-based gradients for choice modelling},
  author={Liu, Yan and Calastri, Chiara and Dekker, Thijs},
  journal={Transportation Planning and Technology},
  pages={1--28},
  year={2025},
  publisher={Taylor \& Francis}
}

@article{buckell2025break,
  title={A break from the norm? Parametric representations of preference heterogeneity for discrete choice models in health},
  author={Buckell, John and Wreford, Alice and Quaife, Matthew and Hancock, Thomas O},
  journal={Medical Decision Making},
  volume={45},
  number={8},
  pages={987--1001},
  year={2025},
  publisher={Sage Publications Sage CA: Los Angeles, CA}
}

@article{van2014specification,
  title={Specification of the utility function in discrete choice experiments},
  author={Van Der Pol, Marjon and Currie, Gillian and Kromm, Seija and Ryan, Mandy},
  journal={Value in Health},
  volume={17},
  number={2},
  pages={297--301},
  year={2014},
  publisher={Elsevier}
}

@article{berndt1974bhhh,
  title={Estimation and Inference in Nonlinear Structural Models},
  author={Berndt, Ernst R. and Hall, Bronwyn H. and Hall, Robert E. and Hausman, Jerry A.},
  journal={Annals of Economic and Social Measurement},
  volume={3},
  number={4},
  pages={653--665},
  year={1974}
}

@book{higham2002accuracy,
  title={Accuracy and Stability of Numerical Algorithms},
  author={Higham, Nicholas J.},
  year={2002},
  publisher={SIAM},
  edition={2nd}
}

@article{mcfadden2000mixed,
  title={Mixed MNL Models for Discrete Response},
  author={McFadden, Daniel and Train, Kenneth},
  journal={Journal of Applied Econometrics},
  volume={15},
  number={5},
  pages={447--470},
  year={2000},
  doi={10.1002/1099-1255(200009/10)15:5<447::AID-JAE570>3.0.CO;2-1}
}

@article{hess2019apollo,
  title={Apollo: A Flexible, Powerful and Customisable Freeware Package for Choice Model Estimation and Application},
  author={Hess, Stephane and Palma, David},
  journal={Journal of Choice Modelling},
  volume={32},
  pages={100170},
  year={2019},
  doi={10.1016/j.jocm.2019.100170}
}

@article{arteaga2022xlogit,
  title={xlogit: An Open-Source Python Package for GPU-Accelerated Estimation of Mixed Logit Models},
  author={Arteaga, Cristian and Park, JeeWoong and Beeramoole, Prithvi Bhat and Paz, Alexander},
  journal={Journal of Choice Modelling},
  volume={42},
  pages={100339},
  year={2022},
  doi={10.1016/j.jocm.2021.100339}
}

@techreport{bierlaire2023biogeme,
  title={A Short Introduction to Biogeme},
  author={Bierlaire, Michel},
  institution={Transport and Mobility Laboratory, ENAC, EPFL},
  number={TRANSP-OR 230620},
  year={2023},
  url={https://transp-or.epfl.ch/documents/technicalReports/Bier23.pdf}
}

@article{greene2003latent,
  title={A Latent Class Model for Discrete Choice Analysis: Contrasts with Mixed Logit},
  author={Greene, William H. and Hensher, David A.},
  journal={Transportation Research Part B: Methodological},
  volume={37},
  number={8},
  pages={681--698},
  year={2003}
}

@article{du2023torchchoice,
  title={Torch-Choice: A PyTorch Package for Large-Scale Choice Modeling with Python},
  author={Du, Tianyu and Kanodia, Ayush and Athey, Susan},
  journal={arXiv preprint arXiv:2304.01906},
  year={2023},
  doi={10.48550/arXiv.2304.01906}
}

@misc{jia2018dissecting,
  title={Dissecting the {NVIDIA} {V}olta {GPU} Architecture via Microbenchmarking},
  author={Jia, Zhe and Maggioni, Marco and Staiger, Benjamin and Scarpazza, Daniele P.},
  howpublished={arXiv preprint arXiv:1804.06826},
  year={2018}
}

@book{kirk2016programming,
  title={Programming Massively Parallel Processors: A Hands-on Approach},
  author={Kirk, David B. and Hwu, Wen-mei W.},
  year={2016},
  edition={3rd},
  publisher={Morgan Kaufmann}
}

@article{fiebig2010gmnl,
  title={The Generalized Multinomial Logit Model: Accounting for Scale and Coefficient Heterogeneity},
  author={Fiebig, Denzil G. and Keane, Michael P. and Louviere, Jordan and Wasi, Nada},
  journal={Marketing Science},
  volume={29},
  number={3},
  pages={393--421},
  year={2010},
  publisher={INFORMS}
}

@book{ben-akiva1985discrete,
  title={Discrete Choice Analysis: Theory and Application to Travel Demand},
  author={Ben-Akiva, Moshe E. and Lerman, Steven R.},
  year={1985},
  publisher={MIT Press},
  address={Cambridge, MA}
}

@article{bhat2003halton,
  title={Simulation Estimation of Mixed Discrete Choice Models Using Randomized and Scrambled {Halton} Sequences},
  author={Bhat, Chandra R.},
  journal={Transportation Research Part B: Methodological},
  volume={37},
  number={9},
  pages={837--855},
  year={2003},
  doi={10.1016/S0191-2615(02)00090-5}
}

@article{hess2006mlhs,
  title={On the Use of a Modified {Latin Hypercube Sampling} ({MLHS}) Method in the Estimation of a Mixed Logit Model for Vehicle Choice},
  author={Hess, Stephane and Train, Kenneth E. and Polak, John W.},
  journal={Transportation Research Part B: Methodological},
  volume={40},
  number={2},
  pages={147--163},
  year={2006},
  doi={10.1016/j.trb.2004.10.005}
}

@article{cherchi2012curse,
  title={A {Monte Carlo} Experiment to Analyze the Curse of Dimensionality in Estimating Random Coefficients Models with a Full Variance--Covariance Matrix},
  author={Cherchi, Elisabetta and Guevara, Cristian Angelo},
  journal={Transportation Research Part B: Methodological},
  volume={46},
  number={2},
  pages={321--332},
  year={2012},
  doi={10.1016/j.trb.2011.10.006}
}

@article{hole2007bootstrap,
  title={A Comparison of Approaches to Estimating Confidence Intervals for Willingness to Pay Measures},
  author={Hole, Arie Rinald},
  journal={Health Economics},
  volume={16},
  number={8},
  pages={827--840},
  year={2007},
  doi={10.1002/hec.1197}
}

@article{mcfadden1986choice,
  title={The Choice Theory Approach to Market Research},
  author={McFadden, Daniel},
  journal={Marketing Science},
  volume={5},
  number={4},
  pages={275--297},
  year={1986},
  doi={10.1287/mksc.5.4.275}
}

@article{lancsar2008conducting,
  title={Conducting Discrete Choice Experiments to Inform Healthcare Decision Making: A User's Guide},
  author={Lancsar, Emily and Louviere, Jordan},
  journal={PharmacoEconomics},
  volume={26},
  number={8},
  pages={661--677},
  year={2008},
  doi={10.2165/00019053-200826080-00004}
}

@article{adamowicz1998stated,
  title={Stated Preference Approaches for Measuring Passive Use Values: Choice Experiments and Contingent Valuation},
  author={Adamowicz, Wiktor and Boxall, Peter and Williams, Michael and Louviere, Jordan},
  journal={American Journal of Agricultural Economics},
  volume={80},
  number={1},
  pages={64--75},
  year={1998},
  doi={10.2307/3180269}
}

@article{czajkowski2019simulation,
  title={Simulation Error in Maximum Likelihood Estimation of Discrete Choice Models},
  author={Czajkowski, Miko{\l}aj and Budzi{\'n}ski, Wiktor},
  journal={Journal of Choice Modelling},
  volume={31},
  pages={73--85},
  year={2019},
  doi={10.1016/j.jocm.2019.04.003}
}

@article{hole2007mixlogit,
  title={Fitting Mixed Logit Models by Using Maximum Simulated Likelihood},
  author={Hole, Arne Risa},
  journal={The Stata Journal},
  volume={7},
  number={3},
  pages={388--401},
  year={2007},
  doi={10.1177/1536867X0700700306}
}

@article{bertoni2026medical,
  title={Medical Brain Drain: The Role of Job Attributes and Individual Traits},
  author={Bertoni, Marco and Chattopadhyay, Debdeep and Gu, Yuanyuan},
  journal={Health Economics},
  year={2026},
  doi={10.1002/hec.70148},
  url={https://doi.org/10.1002/hec.70148}
}
